\documentclass[useAMS,fleqn,usenatbib]{mn2e}
\usepackage{amssymb,amsmath}
\usepackage{mnras_macros}
\usepackage{graphicx}
\usepackage[pdftex]{color}
\def\aap{A\&A}%

\def\aj{AJ}%
\def\apj{ApJ}%
\def\apjl{ApJ}%
\def\apjs{ApJS}%
\def\ao{Appl.~Opt.}%
\def\jcap{J. Cosmology Astropart. Phys.}%
\def\mnras{MNRAS}%
\def\nar{New A Rev.}%
\def\nat{Nature}%
\def\pasj{PASJ}%
\def\physrep{Phys.~Rep.}%
\def\prd{Phys.~Rev.~D}%
\def\procspie{Proc.~SPIE}%
\title[Annihilation and decay of non-spherical dark halos]{DARK MATTER ANNIHILATION AND DECAY FROM NON-SPHERICAL DARK HALOS IN THE GALACTIC DWARF SATELLITES}

\author[K. Hayashi et al.]{Kohei Hayashi$^{1}$\thanks{Contact e-mail: kohei.hayashi@ipmu.jp}, Koji Ichikawa$^{1}$, Shigeki Matsumoto$^{1}$, Masahiro Ibe$^{1,2}$, 
\newauthor Miho N. Ishigaki$^{1}$ and Hajime Sugai$^{1}$
\\
$^{1}$Kavli Institute for the Physics and Mathematics of the Universe (Kavli IPMU, WPI), The University of Tokyo, Chiba 277-8583, Japan\\
$^{2}$Institute for Cosmic Ray Research (ICRR), The University of Tokyo, Chiba 277-8583, Japan\\
}

\date{Accepted 2016 June 15. Received 2016 June 2; in original form 2016 March 24}

\pagerange{\pageref{firstpage}--\pageref{lastpage}}
\pubyear{2016}

\begin{document}
\maketitle
\label{firstpage}

\begin{abstract}
The dwarf spheroidal galaxies~(dSphs) in the Milky Way are the primary targets in the indirect searches for particle dark matter.
To set robust constraints on candidate of dark matter particles, understanding the dark halo structure of these systems is of substantial importance.
In this paper, we first evaluate the astrophysical factors for dark matter annihilation and decay for 24~dSphs, with taking into account a non-spherical dark halo, using generalized axisymmetric mass models based on axisymmetric Jeans equations.
First, from a fitting analysis of the most recent kinematic data available, our axisymmetric mass models are a much better fit than previous spherical ones, thus, our work should be the most realistic and reliable estimator for astrophysical factors. 
Secondly, we find that among analysed dSphs, the ultra-faint dwarf galaxies Triangulum~II and Ursa~Major~II are the most promising but large uncertain targets for dark matter annihilation while the classical dSph Draco is the most robust and detectable target for dark matter decay.
It is also found that the non-sphericity of luminous and dark components influences the estimate of astrophysical factors, even though these factors largely depend on the sample size, the prior range of parameters and the spatial extent of the dark halo.
Moreover, owing to these effects, the constraints on the dark matter annihilation cross-section are more conservative than those of previous spherical works.
These results are important for optimizing and designing dark matter searches in current and future multi-messenger observations by space and ground-based telescopes.
\end{abstract}


%%%%%%%%%%%%%%%%%%%%%%%%%%%%%%%%%%%%%%%%%%%%%%%%%%

%%%%%%%%%%%%%%%%% BODY OF PAPER %%%%%%%%%%%%%%%%%%
%%% Sec.1 %%%%%%%%%%%%%%%%%%%%%%%%%%%%%%%%%%%%%%%%
\section{Introduction}
Owing to the precise data on the cosmic microwave background \citep[e.g.,][]{Kometal2011,Planck2015}, it is well established that around 80 per~cent of the mass density of matter in the Universe consists of non-baryonic matter, known as dark matter, which is formed of still unidentified matter particles. Indirect searches for dark matter annihilations or decays are one of the unique ways of constraining numerous particle candidates for dark matter~\citep[e.g.,][for review]{Gunetal1978,Ber2012}.
In particular, the indirect detections utilizing $\gamma$-ray or X-rays emissions are suitable for limiting the heavier dark matter candidates of $\sim 0.1$ -- 1~TeV, such as weakly interacting massive particles (WIMPs). 
To implement such an indirect search effectively, we should look at the densest dark matter regions, like the Galactic centre, dwarf spheroidal galaxies~(dSphs) and galaxy clusters, because these regions could be the source of detectable high-energy photons produced in dark matter annihilation or decay events, despite being undetected so far.

The Galactic dSph satellites, which we focus on in this work, are ideal sites for constraining particle candidates of dark matter through indirect searches for their annihilations and decays~\citep[e.g.,][]{Lake1990,Str2013,Wal2013} because of their proximity, high dynamical mass-to-light ratios ($M/L\sim10 -1000$) and lack of astrophysical contaminating $\gamma$-ray sources~\citep{Giletal2007,McC2012}.
Thanks to Sloan Digital Sky Survey~\citep[SDSS,][]{Yoretal2000}, such dwarf satellites and more dark matter-dominated systems have been discovered by some observational studies~\citep{Wiletal2005,Beletal2007,Beletal2008,Beletal2009,Beletal2010,Zucetal2006a,Zucetal2006b}.  
More recently, using the Dark Energy Survey~\citep{Fla2005} and Pan-STARRS~1~3$\pi$ survey~\citep{Kaietal2002}, 12 faint dwarf galaxies were discovered~\citep{Becetal2015,Laeetal2015a,Laeetal2015b}.
Therefore, the number of promising targets of indirect searches for dark matter particles have been dramatically increased by the advanced imaging surveys over the past decade, and further, new targets should be found by the very deep and wide photometric surveys of large telescopes such as Subaru/Hyper Suprime Cam~\citep[HSC,][]{Miyetal2012} and the Large Synoptic Survey Telescope~\citep[LSST,][]{Tysetal2003} in the near future.  

Previous studies of the indirect detections of dSphs obtained several upper limits on the thermally averaged dark matter annihilation cross-section $\langle\sigma v\rangle$ using $\gamma$-ray observations.
By observing and stacking data of five dSphs, the HESS collaboration~\citep{Abretal2014} inferred that the upper limits of $\langle\sigma v\rangle$~$<3.9\times10^{-24}$~cm$^3$~s$^{-1}$ (95 per~cent confidence level) for 1 -- 2~TeV dark matter particle masses.
The MAGIC collaboration~\citep{Aleetal2014} reported that using 160~h Segue~I data, their strongest limit for the self-annihilation cross section is $\langle\sigma v\rangle$~$<1.2\times10^{-24}$~cm$^3$~s$^{-1}$ (95\% per~cent confidence level) for a $\sim$500-GeV dark matter particle annihilating into the $\tau^{+}\tau^{-}$ channel.
Until now, for the heavier dark matter mass range, these upper limits are the strictest limits based on observations by ground-based air Cherenkov telescopes. 
At lower energies, the Large Area Telescope of the {\it Fermi} satellite~(Fermi-LAT) was able to provide $\langle\sigma v\rangle$~$<1.2\times10^{-26}$~cm$^3$~s$^{-1}$ below ${\cal O}(100)$~GeV particle masses annihilating into $b\bar{b}$ based on the   observational results of several dSphs~\citep{Chaetal2011,CS2012,Geretal2015a,Acketal2015}.
Most recently, \citet{Magic2016} performed a joint analysis of $\gamma$-ray data from the MAGIC and Fermi-LAT to obtain severer limits on dark matter candidates, and they improved on the previous upper limits by up to factor of $\sim2$ in the low mass range.

As aforementioned, we highlight that the dSphs in the Milky Way are the most promising targets for the indirect detection of dark matter through $\gamma$-ray observations. 
These studies, however, largely rely on the astrophysical factor (the so-called $J$ and $D$ factors, as we discuss later) and thus, require a very meticulous and optimal evaluation of dark halo structures in the dSphs. To investigate dark matter distributions (and their uncertainties) in the dSphs, most studies have constructed dynamical mass models based on Jeans equations and applied them to the line-of-sight velocity data of dSph member stars. In fitting dynamical mass models to kinematic data, most authors assume that the density distributions of both the member stars and dark matter are spherically symmetric and utilize these profiles to solve the spherical Jeans equation to obtain the projected velocity dispersion in the line of sight~\citep[e.g.,][]{Chaetal2011,Geretal2015a,Geretal2015,Bonetal2015a,Bonetal2015b}.
For instance, \citet[][hereafter GS15]{Geretal2015} evaluated the expected astrophysical factors for 20 Milky Way dSphs using a spherical Jeans analysis of the most recent stellar line-of-sight velocity data available, and suggested that the ultra-faint dwarf (UFD) galaxies Ursa~Major~II and Segue~I are the most attractive targets among the analysed dSphs, even though these galaxies have large uncertainties in their astrophysical factors due to the small spectroscopic data sample.

However, there are various embarrassing but crucial systematic uncertainties of dark halo evaluations in the above dynamical analysis: the free parameter prior bias, stellar velocity anisotropy, unresolved binary stars, the size of the dark halo, non-sphericity and foreground contamination. These effects are not negligible even for classical dSphs and UFD galaxies. For the effects of contamination, as an example, \citet{Bonetal2015a}~investigated how estimations of $J$-factors are sensitive to foreground contamination of the stellar-kinematic data of the UFD galaxy Segue~I. They found that Milky Way halo stars as contamination can induce the astrophysical factors to be overestimated by orders of magnitude. 
Besides, classical dSphs, which have relatively large data sample, are also affected by foreground contamination, as reported by~\citet{Ichetal2016}.
They inspected this systematic bias in dark halo estimation using a new likelihood function that includes the foreground effect and then concluded that although the fraction of misclassified contamination is less than 5~per~cent, the fluctuation at most is $\delta\log_{10}J\sim0.1$, even with a ${\cal O}(1000)$ data sample.
Thus, several works have begun to inspect the influence of these systematics on astrophysical factor estimations.

Meanwhile, the non-sphericity of actual stellar systems and their dark haloes is not taken much into consideration with respect to the evaluation of astrophysical factors, even though some groups have examined this effect using a mock data catalogue~\citep{Bonetal2015a}.
As mentioned above, most previous studies have treated the dSphs and their dark haloes as spherical systems for simplicity, despite that the observed luminous parts of the dSphs are actually non-spherical~\citep[e.g.,][]{IH1995,McC2012} and the concordance $\Lambda$-cold dark matter models predict non-spherical virialized dark haloes~\citep[e.g.,][]{JS2002,Kuhetal2007,Veretal2014}.
Thus, we should address non-spherical mass models for dSph galaxies rather than spherical mass models to obtain more reliable and realistic limits on the properties of the dark halo.
\citet[][]{HC2012}~constructed axisymmetric mass models based on axisymmetric Jeans equations and applied these models to line-of-sight velocity dispersion profiles of six classical Milky Way dSphs. Furthermore, \citet[][hereafter HC15]{HC2015b} revisited the dynamical analysis of the dSphs based on revised axisymmetric mass models from~\citet{HC2012}.
Both axisymmetric studies concluded that most of the dSphs have very oblate and flattened dark haloes. Although we here follow the axisymmetric mass models developed by HC15 and apply these models to the most recent kinematic data of the 17 UFD galaxies as well as seven classical dSphs, we adopt an unbinned analysis for the comparison between data and models, unlike previous axisymmetric works. Since the dSphs, especially UFD galaxies, do not have not a large amount of kinematic data due to the very faint systems, an unbinned analysis is the more robust method for constraining dark halo parameters rather than a binned analysis. Moreover, the important point in this work is that we first calculate the astrophysical factors of the Milky Way dSphs by considering a non-spherical dark halo.
 
This paper is organized as follows. 
In Section~2, we introduce the astrophysical factors $J$ and $D$ calculated from axisymmetric dark matter distributions.
In Section~3, we explain the axisymmetric models for the density profiles of stellar and dark halo components based on a Jeans analysis.
In Section~4 and 5, we briefly describe the photometric and spectroscopic data for seven classical and 17 UFD spheroidal galaxies in the Milky Way, and introduce the method of fitting to the data and likelihood function we adopted, respectively.
In Section~6, we present the results of the fitting analysis and the astrophysical factor values, and compare them with previous works. We also estimate the sensitivity line of the dark matter annihilation cross-section with respect to dark matter particle mass using our estimated dark halo properties.
Finally, our conclusions are presented in Section~7.

%%% Sec.2 %%%%%%%%%%%%%%%%%%%%%%%%%%%%%%%%%%%%%%%%
% Figure 1
\begin{figure}
	\begin{center}
	\includegraphics[width=\columnwidth]{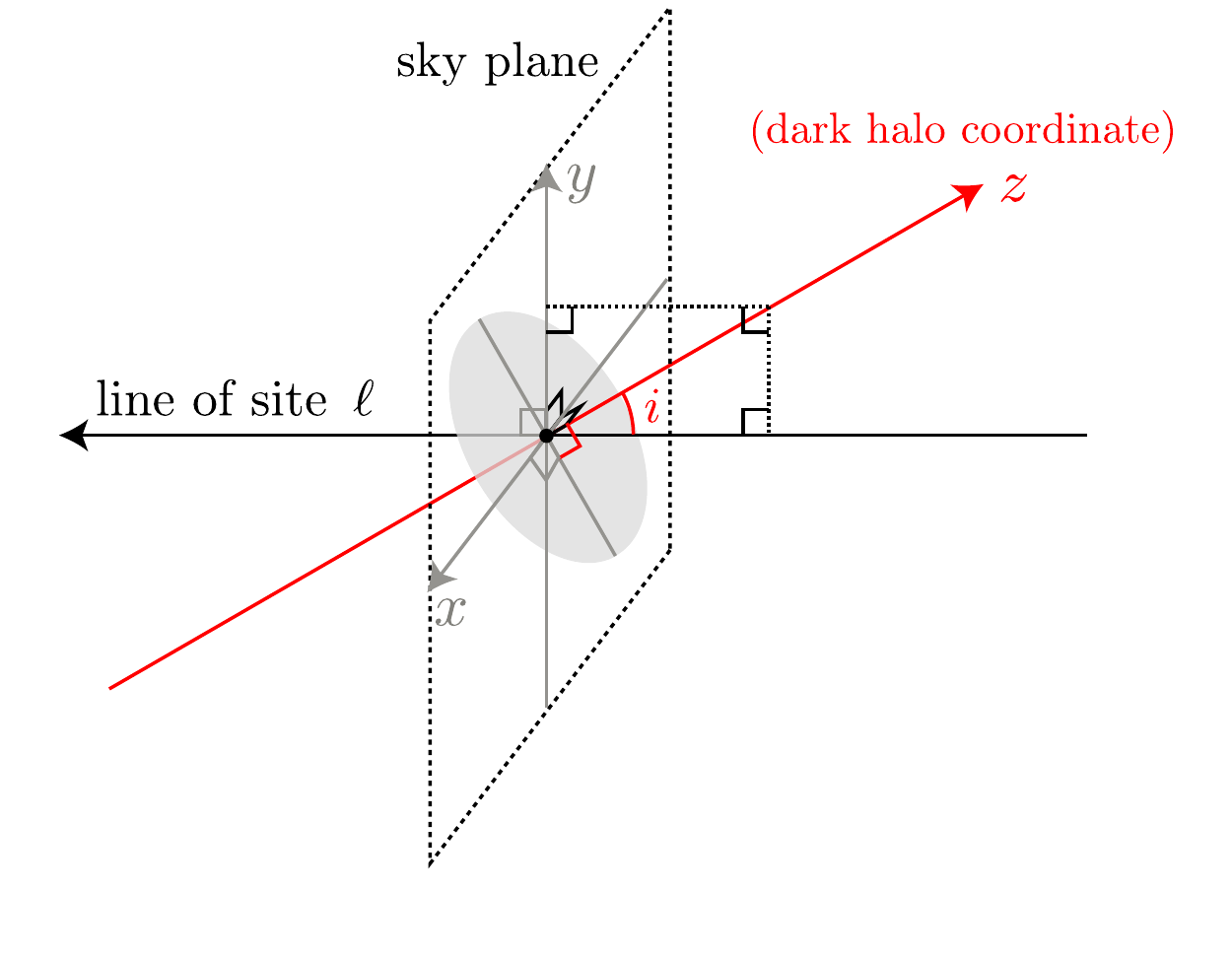}
	\end{center}
    \caption{Scheme of the relation between the coordinates of the axisymmetric dark halo and the sky plane with respect to the line of sight. The grey spheroidal denotes the axisymmetric dark halo in cylindrical coordinates $(R,\phi,z)$. $(x,y)$ and $\ell$ are the coordinates for the sky plane and line of sight, respectively. We define that the dark halo coordinate $\phi=0$ corresponds to the $x$-axis. The $z$-axis is identical to the $y$-axis for $i=90^{\circ}$, i.e. edge-on.}
    \label{fig:integral}
\end{figure}

\section{Astronomical factor for annihilation and decay}
The $\gamma$-ray flux from dark matter annihilation and decay in a dSph, measured over a solid angle $\Delta\Omega$, are estimated by two important factors.
The first factor is based on the microscopic physics of dark matter while the second reflects its distribution on astronomical scales. The former factor is related to the particle mass of dark matter, the velocity-averaged annihilation cross-section (or decay rate for each volume, for the other factor), the number of photons per energy produced by annihilation (decay), and the annihilation (decay) branching fraction, and thus, it is largely dependent on the particle’s physics properties. What we focus on here is given by 
\begin{eqnarray}
J &=& \int_{\Delta\Omega}\int_{\rm los}d\ell d\Omega \hspace{1mm}\rho^2(\ell,\Omega) \hspace{3.5mm} {\rm [annihilation]}, \label{eq:J}\\
D &=& \int_{\Delta\Omega}\int_{\rm los}d\ell d\Omega \hspace{1mm}\rho(\ell,\Omega) \hspace{4.9mm} {\rm [decay]},
\label{eq:D}
\end{eqnarray}
which are called $J$~factor and $D$~factor~\citep[e.g.,][]{Gunetal1978,Beretal1998,Geretal2015}.
These factors correspond to the line-of-sight integrated dark matter density squared for annihilation and the dark matter density for decay, respectively, within solid angle $\Delta\Omega$. 
The goal of dark matter indirect detection is to obtain severe limits on the nature of the dark matter particles, such as mass and annihilation cross-section, using accurate determinations of the $J$ and $D$~factors along with the observation of $\gamma$-ray.

When we calculate these astrophysical factors, we consider axisymmetric dark matter distributions and thus, these factors are functions of $(x,y)$, which are the projected coordinates on the sky plane, so that the integral variables in equations (\ref{eq:J}) and (\ref{eq:D}) are able to transform $d\ell d\Omega$ into $D^{-2}dxdyd\ell$, where $D$ is the distance from the observer to the center of the dSph.
We also consider the effects of the inclination of the dark matter distributions on these factors.
From Fig.~\ref{fig:integral}, we define the inclination angle $i$ (which is described later) between $z$ and the line of sight, and then the dark halo coordinates $(R,\phi,z)$ are related to the ones in the sky $(x,y)$ and $\ell$ by
%Jacobian matrix
\begin{eqnarray}
\left( 
\begin{array}{ccc}
1 & 0 & 0 \\
0 & -\sin i & -\cos i\\
0 & -\cos i & \sin i \\
\end{array} 
\right)
\left( 
\begin{array}{c}
x \\
\ell\\
y  \\
\end{array} 
\right)
=\left( 
\begin{array}{c}
R\cos\phi \\
R\sin\phi\\
z  \\
\end{array} 
\right).
\end{eqnarray}
Since the azimuthal component can be written as $\phi=\cos^{-1}(x/R)$, de-projected coordinates $(R,z)$ can be described geometrically by the projected coordinates and the line-of-sight distance as follows:
\begin{eqnarray}
R^2 &=& x^2 + (y\cos i+\ell\sin i)^2,\\
z^2 &=& (y\sin i -\ell\cos i)^2.
\label{GEO}
\end{eqnarray}

From equations (\ref{eq:J}) and (\ref{eq:D}), the extent of the dark matter distribution is required in calculating the $J$ and $D$~factors. However, it is impossible for us to identify the edge of a dark matter halo using current observational data and dynamical analysis. In this work, we adopt the outermost observed member star $(x_{\rm max},y_{\rm max})$ as the edge of the dark matter halo because there is no clear criterion to define the size of a dark matter halo. As defined by $(x,y)$ above, the value of the outermost star is {\it not} the de-projected distance~(estimated by GS15) {\it but} the line-of-sight projected distance from the center of the dSph.
Even if this underestimates the $J$ and $D$~factor, we regard it as the most conservative dark matter halo size and use it in this paper. Thus, our estimation of astrophysical factors is suitable for placing a conservative and robust constraint on the dark matter annihilation cross-section.

%%% Sec.3 %%%%%%%%%%%%%%%%%%%%%%%%%%%%%%%%%%%%%%%%%%%%%%%%
\section{Models and Jeans analysis}
The dark matter density distribution of dSphs is the key to evaluating their $J$ and $D$~factors.
There are various approaches to estimating the dark halo profile from stellar luminous and kinematic data, such as the developing distribution function, the Schwarzschild and made- to-measure methods, the projected virial theorem as well as Jeans analysis~\citep[see, e.g.,][for review]{BHB2013}.
In this work, we adopt the latter, in particular, using axisymmetric mass models based on axisymmetric Jeans equations to describe the internal structures of the non-spherical dark haloes and stellar systems in dSphs. Here we apply the generalized axisymmetric mass models constructed by HC15, where they took into account the effects of the velocity anisotropy of tracer stars on dark halo parameters, in particular, the shape of the dark halo, because of a strong degeneracy between the velocity anisotropy and the axial ratio of the dark halo~(see Figure~9 in \citealt[][]{BHB2013}; also see Figure~12 in \citealt[][]{HC2015b}).

%%% Sec.3.1 %%%%%%%%%%%%%%%%%%%%%%%%%%%%%%%%%%%%%%%%%%%%%%%%
\subsection{Axisymmetric Jeans equations}
Since a dSph has a relaxation time much longer than its lifetime, these galaxies may be considered as collisionless systems. The stellar spatial and velocity distributions of such dynamical systems are described by a phase-space distribution function. When the system is in dynamical equilibrium and collisionless under the gravitational smooth potential, the distribution function obeys the steady-state collisionless Boltzmann equation~\citep{BT2008}.
In the axisymmetric case, a specific but well-studied consideration is to suppose that the distribution function is of the form $f(E,L_z)$\footnote{$E$ and $L_z$ denote the binding energy and angular momentum component towards the symmetry axis, respectively.}, in which case the mixed moments vanish and the velocity second moments $(\overline{v^2_R}, \overline{v^2_{\phi}}, \overline{v^2_z})$ in cylindrical coordinates $(R,\phi,z)$ are shown as $\overline{v^2_R}=\overline{v^2_z}$ \citep[e.g.,][]{BT2008}.  

More generally, since the velocity second moments in the radial $R$ and vertical $z$ directions are not identical, \citet{Cap2008} relaxed $\overline{v^2_R}=\overline{v^2_z}$ and considered a velocity anisotropy parameter, $\beta_z=1-\overline{v^2_z}/\overline{v^2_R}$, where $\beta_z$ is assumed to be constant. His formalism was widely used for the kinematic structure of early-type galaxies \citep[e.g.,][]{Scoetal2009,Setetal2010,Labetal2012}. Moreover, HC15 adopted his anisotropic solution to the kinematical data of the dSphs in Milky Way and Andromeda.
This assumption is passably supported by pure dark matter simulations reported by \cite{Veretal2014}.
In addition, we assume that a dSph does not rotate and thus, the second moment is equal to the velocity dispersion because their line-of-sight velocity gradients are so small. Moreover, the origin of their small velocity gradients can be understood by projection effects in many cases~\citep[e.g.,][]{Batetal2008,Batetal2011,Kocetal2007a,Matetal2008,Waletal2008}.
Even if these galaxies have rotation velocity, the velocity to dispersion ratio, $v/\sigma$, is significantly low (i.e., a dSph is a dispersion supported system).
Therefore, the influence of rotating would not be so large.

Under these assumptions, the second-order axisymmetric Jeans equations are derived from taking the velocity moments of the collisionless Boltzmann equation:
\begin{eqnarray}
\overline{v^2_z} &=&  \frac{1}{\nu(R,z)}\int^{\infty}_z \nu\frac{\partial \Phi}{\partial z}dz,
\label{AGEb03}\\
\overline{v^2_{\phi}} &=& \frac{1}{1-\beta_z} \Biggl[ \overline{v^2_z} + \frac{R}{\nu}\frac{\partial(\nu\overline{v^2_z})}{\partial R} \Biggr] + 
R \frac{\partial \Phi}{\partial R}.
\label{AGEb04}
\end{eqnarray}
where $\nu$ is the three-dimensional stellar density and $\Phi$ is gravitational potential dominated by dark matter. These velocity second moments are provided by the second moments that separate into the contribution of ordered and random motions, as defined by~$\overline{v^2}=\sigma^2+\overline{v}^2$.
We also assume for simplicity that the density of the tracer stars has the same orientation and symmetry as that of the dark halo.

These equations indicate the intrinsic velocity second moments, hence they should be integrated along the line of sight to be compared with the observable velocity second moments for the dSph.
Thus we derive these from $\overline{v^2_R}$ $(=(1-\beta_z)^{-1}\overline{v^2_z})$, $\overline{v^2_z}$ and $\overline{v^2_{\phi}}$, considering the inclination of the dSph with respect to the observer, following the method given in~\citep{TT2006,HC2012}.
First, for $\overline {v^2_R}$ and $\overline{v^2_{\phi}}$, we project these to the plane parallel to the galactic plane as 
\begin{equation}
\overline{v^2_{\ast}} = \overline{v^2_{\phi}}\frac{x^2}{R^2} + \overline{v^2_{R}}\Bigl(1-\frac{x^2}{R^2} \Bigr),
\label{LOS1}
\end{equation}
where $x$ is the projected coordinate. 
Secondly, using the inclination $\theta$ as the angle between the galactic plane and the line of sight, we project $\overline{v^2_{z}}$ and $\overline{v^2_{\ast}}$ to the line-of-sight:
\begin{equation}
\overline{v^2_{\ell}} = \overline{v^2_{\ast}} \cos^2\theta + \overline{v^2_{z}}\sin^2\theta.
 \label{LOS2}
\end{equation}
Finally, we calculate the averaged $\overline{v^2_{\ell}}$ along the line of sight, so that we compare the theoretical velocity second moments with the observed data.
The line-of-sight velocity second moment\footnote{As we solve the line-of-sight velocity second moments, we compute three dimensional integrals, namely the integral in equations~(\ref{AGEb03}), (\ref{LOS3}) and (\ref{force}), using Gauss-Legendre quadrature. For the validity of our calculations, we compared the line-of-sight velocity second moments calculated from our numerical Jeans solution with those from the analytical solution derived by~\citet{Evans1993}, and confirmed that our calculations are quite identical to analytic cases.} is 
\begin{equation} 
\overline{v^2_{\rm los}}(x,y) = \frac{1}{I(x,y)}\int^{\infty}_{-\infty} \nu(R,z)\overline{v^2_{\ell}}(R,z)d\ell,
\label{LOS3}
\end{equation}
where $I(x,y)$ denotes the surface density profile integrated $\nu(R,z)$ along the line of sight $\ell$.

In this work, we assume that member stars in the dSph are distributed according to an {\it oblate}\footnote{Here, we suppose that the stellar distribution of a dSph has an oblate shape only. This is because HC15 concluded that most oblate stellar distributions are a much better fit than prolate ones.} Plummer profile \citep{Plu1911} generalized to an axisymmetric form:
\begin{equation}
\nu(R,z) = \frac{3L}{4\pi b^3_{\ast}} \Bigl[1+\frac{m^2_{\ast}}{b^2_{\ast}}\Bigr]^{-5/2},
\label{3DPlu}
\end{equation}
where $m^2_{\ast} = R^2 + z^2/q^2$, so $\nu$ is constant on spheroidal shells with axial ratio $q$, and $L$ and $b_{\ast}$ are the total luminosity and half light radius, respectively.
This spatial density can be used analytically to calculate the surface density,
\begin{equation}
I(x,y)=\frac{L}{\pi b^2_{\ast}} \frac{1}{(1+m^{\prime 2}_{\ast}/b^2_{\ast})^{2}},
\label{2DPlu}
\end{equation}
where $m^{\prime 2}_{\ast}=x^2+y^2/q^{\prime 2}$, $q^{\prime}$ is the projected axial ratio and $(x,y)$ are the sky plane coordinates aligned with the projected major and minor axes, respectively. Projected axial ratio is related to the intrinsic one and inclination angle $i$~$(=90^{\circ}-\theta)$, as $q^{\prime 2}=\cos^2i + q^2\sin^2i$, where $i=90^{\circ}$ when a galaxy is seen in edge-on~(see Fig~\ref{fig:integral}). 
The intrinsic axial ratio can be derived from $q=\sqrt{q^{\prime 2}-\cos^2i}/\sin i$, and thus, this angle is restricted by $q^{\prime 2} -\cos^2i>0$.

%%% Table 1 %%%
%%% Table 1 %%%
\begin{table*}
	\centering
	\caption{Observational data set for MW dSph satellites.}
	\label{table1}
	\begin{tabular}{ccccccccc} % four columns, alignment for each
		\hline\hline
Object & $N_{\rm sample}$ & RA(J2000) & DEC(J2000)   &$M_V$  & $D_{\odot}$ & $b_{\ast}$ &  $q^{\prime}$ &   Ref.$^{a}$ \\
            &                                 & [hh:mm:ss]   &  [dd:mm:ss]     &              &      [kpc]         &        [pc]              &    (axial ratio)  &  \\
		\hline
		{\bf Classical dwarfs}     &&&&&&&&\\
		Carina          & 776    & 06:41:36.7   & $-$50:57:58     & $-9.1  \pm 0.5 $  & $106\pm 6$  & $250\pm  39$       & $0.67\pm0.05$   & 1,6 \\
		Fornax          & 2523   & 02:39:59.3   & $-$34:26:57     & $-13.4 \pm 0.3 $  & $147\pm12$  & $710\pm  77$       & $0.70\pm0.01$   & 1,6 \\
		Sculptor        & 1360   & 01:00:09.4   & $-$33:42:33     & $-11.1 \pm 0.5 $  & $ 86\pm 6$  & $283\pm  45$       & $0.68\pm0.03$   & 1,6 \\
		Sextans         & 445    & 10:13:03.0   & $-$01:36:53     & $-9.3  \pm 0.5 $  & $ 86\pm 4$  & $695\pm  44$       & $0.65\pm0.05$   & 1,6 \\
		Draco           & 468    & 17:20:12.4   & $+$57:54:55     & $-8.8  \pm 0.3 $  & $ 76\pm 6$  & $221\pm  19$       & $0.69\pm0.02$   & 1,7 \\
		Leo~I           & 328    & 10:08:28.1   & $+$12:18:23     & $-12.0 \pm 0.3 $  & $254\pm15$  & $251\pm  27$       & $0.79\pm0.03$   & 1,8 \\
		Leo~II          & 200    & 11:13:28.8   & $+$22:09:06     & $-9.8  \pm 0.3 $  & $233\pm14$  & $176\pm  42$       & $0.87\pm0.05$   & 1,9 \\
		%Ursa Minor & No data \\
		{\bf Ultra faint dwarfs}    &&&&&&&&\\
		Segue~1        &    73  & 10:07:04.0   & $+$16:04:55      & $-1.5 \pm 0.8 $   &$ 32\pm 6$        & $29^{+8}_{-5}$     & $0.53\pm0.10$   & 1,10\\
		Segue~2        &    24  & 02:19:16.0   & $+$20:10:31      & $-2.5 \pm 0.3 $   &$ 35\pm 2$        & $35\pm3$           & $0.85\pm0.13$   & 1,11\\
		Bo\"otes I     &    37  & 14:00:06.0   & $+$14:30:00      & $-6.3 \pm 0.2 $   &$ 66\pm 2$        & $242\pm21$         & $0.61\pm0.06$   & 1,12\\
		Hercules       &    18  & 16:31:02.0   & $+$12:47:30      & $-6.6 \pm 0.4 $   &$132\pm12$        & $330^{+75}_{-52}$  & $0.32\pm0.08$   & 1,13\\
 Coma Berenices        &    59  & 12:26:59.0   & $+$23:54:15      & $-3.7 \pm 0.6 $   &$ 44\pm 4$        & $64\pm7$           & $0.62\pm0.14$   & 1,14\\ 
 Canes~Venatici~I      &   214  & 13:28:03.5   & $+$33:33:21      & $-7.9 \pm 0.5 $   &$224^{+22}_{-20}$ & $554\pm63$         & $0.61\pm0.03$   & 1,14\\
 Canes~Venatici~II     &    25  & 12:57:10.0   & $+$34:19:15      & $-4.8 \pm 0.6 $   &$151^{+15}_{-13}$ & $132\pm16$         & $0.48\pm0.11$   & 1,14\\
		Leo~IV         &    18  & 11:32:57.0   & $-$00:32:00      & $-5.1 \pm 0.6 $   &$158^{+15}_{-14}$ & $152\pm17$         & $0.51\pm0.11$   & 1,14\\
		Leo~V          &     5  & 11:31:09.6   & $+$02:13:12      & $-5.2 \pm 0.4 $   &$178\pm10$        & $135\pm32$         & $0.50\pm0.15$   & 1,15\\
		Leo~T          &    19  & 09:34:53.4   & $+$17:03:05      & $-7.1 \pm 0.3 $   &$417^{+20}_{-19}$ & $170\pm15$         & $\sim1.00$      & 1,14\\
%	Willman~I        &       0  & 09:34:53.4   & $+$17:03:05      & $-7.1 \pm 0.3 $   &$417^{+20}_{-19}$ & $170\pm15$         & \\
		Ursa Major~I   &    39  & 10:34:52.8   & $+$51:55:12      & $-5.6 \pm 0.6 $   &$106^{+9}_{-8}$   & $308\pm32$         & $0.20\pm0.04$   & 1,14\\
	   Ursa Major~II   &    20  & 08:51:30.0   & $+$63:07:48      & $-3.8 \pm 0.6 $   &$ 32^{+5}_{-4}$   & $127\pm21$         & $0.37\pm0.05$   & 1,14\\
		Reticulum~II   &    25  & 03:35:42.1   & $-$54:02:57      & $-2.7 \pm 0.1 $   &$32\pm3$          & $32^{+2}_{-1}$     & $0.41\pm0.03$   & 2,16\\
		Draco~II       &     9  & 15:52:47.6   & $+$64:33:55      & $-2.9 \pm 0.8 $   &$20\pm3$          & $19^{+8}_{-6}$     & $0.76^{+0.27}_{-0.24}$   & 3,17\\ 
		Triangulum~II  &    13  & 02:13:17.4   & $+$36:10:42      & $-1.8 \pm 0.5 $   &$30\pm2$          & $34^{+9}_{-8}$     & $0.79^{+0.17}_{-0.21}$   & 4,18\\
		Hydra~II       &    13  & 12:21:42.1   & $-$31:59:07      & $-4.8 \pm 0.3 $   &$134\pm10$        & $68\pm11$          & $0.99^{+0.01}_{-0.19}$   & 5,19\\
		Pisces~II      &     7  & 22:58:31.0   & $+$05:57:09      & $-5.0 \pm 0.5 $   &$\sim180$         & $\sim60$           & $0.60\pm0.10$            & 1,19\\
	\hline
	\end{tabular}
\begin{flushleft}
$^{a}$References: (1)~\citet{McC2012}; (2)~\citet{Becetal2015}; (3)~\citet{Laeetal2015b}; (4)~\citet{Laeetal2015a}; (5)~\citet{Maretal2015b}; (6)~\citet{Waletal2009a}; (7)~\citet{Waletal2015}; (8)~\citet{Matetal2008}; (9)~\citet{Kocetal2007}; (10)~\citet{Simetal2011}; (11)~\citet{Kiretal2013}; (12)~\citet{Kopetal2011}; (13)~\citet{Adeetal2009}; (14)~\citet{SG2007}; (15)~\citet{Waletal2009c}; (16)~\citet{Simetal2015}; (17)~\citet{Maretal2015a}; (18)~\citet{Maretal2016}; (19)~\citet{Kiretal2015};
\end{flushleft}
\end{table*}

%%% Sec.3.2 %%%%%%%%%%%%%%%%%%%%%%%%%%%%%%%%%%%%%%%%%%%%%%%%
\subsection{Dark Matter Halo Model}
According to various previous papers \citep[e.g.,][]{Giletal2007,Stretal2010,HC2012}, the dark matter density profile in dSphs, especially the inner slope, is not necessarily the cusped profiles predicted by $\Lambda$CDM based $N$-body simulations.
Therefore, we do not confine ourselves to cosmological motivated profiles such as the Navarro-Frenk-White (NFW) \citep{NFW1996,NFW1997} or Einasto profiles~\citep[e.g.,][]{Navetal2010}. 
Following~HC15, we here consider the following profiles to describe the non-spherical dark matter haloes of dSphs:
\begin{eqnarray}
&& \rho(R,z) = \rho_0 \Bigl(\frac{m}{b_{\rm halo}} \Bigr)^{\alpha}\Bigl[1+\Bigl(\frac{m}{b_{\rm halo}} \Bigr)^2 \Bigr]^{-(\alpha+3)/2},
 \label{DMH} \\
&& m^2=R^2+z^2/Q^2,
\label{DMH2}
 \end{eqnarray}
where the four free parameters are the normalized density $\rho_0$ and a transition radius $b_{\rm halo}$, between the inner and outer slopes of the spatial distribution, the inner slope $\alpha$, and $Q$, which is a constant axial ratio of dark matter haloes.
For simplicity, we assume that the slope of dark matter profiles in the outer parts consistently has $\rho\propto r^{-3}$, so that $\alpha=-1$,which is well known as a NFW cuspy profile, while $\alpha=0$ corresponds with a Burkert cored profile \citep{Bur1995}.
The vantage point of this assumed profile is that the form of equations (\ref{DMH}) and (\ref{DMH2}) allows as to calculate the gravitational force in a straightforward way\citep{vanetal1994,BT2008,HC2012}. 
By using a new variable of integration $\tau$, the gravitation force can be obtained by the one-dimensional integration:
\begin{equation}
{\bf g} = -\nabla\Phi = -\pi GQa_0\int^{\infty}_{0} d\tau\frac{\rho(\tilde{m}^2)\nabla \tilde{m}^2}{(\tau+a^2_0)\sqrt{\tau+Q^2a^2_0}},
\label{force}
\end{equation} 
where $\tilde{m}^2$ is defined by
\begin{equation}
\frac{\tilde{m}^2}{a^2_0} = \frac{R^2}{\tau+a^2_0} + \frac{z^2}{\tau+Q^2a^2_0}.
\end{equation} 

In this work, we adopt six parameters $(Q,b_{\rm halo},\rho_0,\beta_z,\alpha,i)$ to be determined by fitting to the observed line-of-sight velocity distribution for each dSph.

%%% Sec.4 %%%%%%%%%%%%%%%%%%%%%%%%%%%%%%%%%%%%%%%%%%%%%%%%
\section{Dwarf spheroidal galaxy data}
In this section, we briefly present the photometric and kinematic data of the member stars in the seven classical and 17 UFD galaxies in the Milky Way for the application of our axisymmetric mass models.

The observed properties of these 24 dSphs are tabulated in Table~\ref{table1}: the number of kinematic sample stars identified as member stars, the central sky coordinates, $V$-band absolute magnitude, distance from the Sun, projected half-light radius, projected stellar axial ratio and the references.
We adopted these fundamental data (except for $N_{\rm sample}$) of classical dSphs from \citet[][see references to original papers therein]{McC2012}, and the data of UFD galaxies from \citet{McC2012} and/or discovery papers for satellites discovered after that work~\citep{Becetal2015,Laeetal2015a,Laeetal2015b,Maretal2015b}.

For the stellar-kinematic data of their member stars, we use the latest data as follows.
For the classicals, we utilize stellar-kinematic data published by \citet{Waletal2015} for Draco, by \citet{Matetal2008} for Leo~I, by \citet{Kocetal2007} for Leo~II, and by \citet{Waletal2009a,Waletal2009b} for Carina, Fornax, Sculptor, and Sextans.
For Coma~Berenices, Canes~Venatici~I, Canes~Venatici~II, Leo~IV, Leo~T, Ursa~Major~I and Ursa~Major~II, we use the velocity data of \citet{SG2007}, who observed them using Keck/Deimos and did data reductions, kindly provided by Josh Simon (private communication).
For Segue~I,Segue~II, Bo\"otes~I, Hercules and Leo~V, we adopt data from \citet{Simetal2011}, \citet{Kiretal2013}, \citet{Kopetal2011}, \citet{Adeetal2009} and \citet{Waletal2009c}, respectively.
The $J$~factors of the above dSphs were estimated by GS15, \citet{Bonetal2015a}, and some related papers.
Thus, we can compare $J$ values estimated by spherical and axisymmetric mass models, as we discuss later. 
In addition to the above, we apply our mass models to recently discovered and spectroscopically observed UFD galaxies and use velocity data published by \citet{Simetal2015} for Reticulum~II, by \citet{Maretal2015a} for Draco~II, by \citet{Maretal2016} for Triangulum~II, by \citet{Kiretal2015} for Hydra~II and Pisces~II.

The methods to identify membership stars and exclude fore- ground contamination (i.e. the Galactic halo stars) depend on the above cited papers. For instance, the member stars for Carina, Fornax, Sculptor and Sextans are identified by the expectation-maximization algorithm described in \citet{Waletal2009b}.
For Draco, \citet{Waletal2015} implemented Bayesian method for the resolved stellar spectra in distinguishing the member stars from contamination.
For the other dSphs, the method for evaluating membership is basically a combination of photometric (e.g. the location in a colour-magnitude diagram and spatial position) and spectroscopic (e.g. velocity distribution, metallicity and Na I equivalent width) selections. We note that removing contamination from the data sample is important in determining the dark halo properties accurately. 
In particular, the estimations of $J$ and $D$ factors of UFD galaxies are sensitive to interlopers due to a lack of data volume \citep[][hereafter B15]{Bonetal2015b}.
However, for the present work, we are concerned primarily with systematic differences of $J$ ($D$) values calculated from between spherical and non-spherical mass models.
Therefore, we do not take into account the membership probability of the kinematic sample for each dSph when we define the likelihood function of Markov-chain Monte Carlo~(MCMC) analysis~\citep[e.g. B15 and][]{Ichetal2016}. In other words, we consider that the kinematic data of each dSph that we use here are surely their member stars. To evaluate the $J$- and $D$-factor values more properly, we will combine axisymmetric mass models with a probability of membership, and will present new results for astrophysical factors in a forthcoming paper.

% Figure 2
\begin{figure*}
	\includegraphics[width=\textwidth]{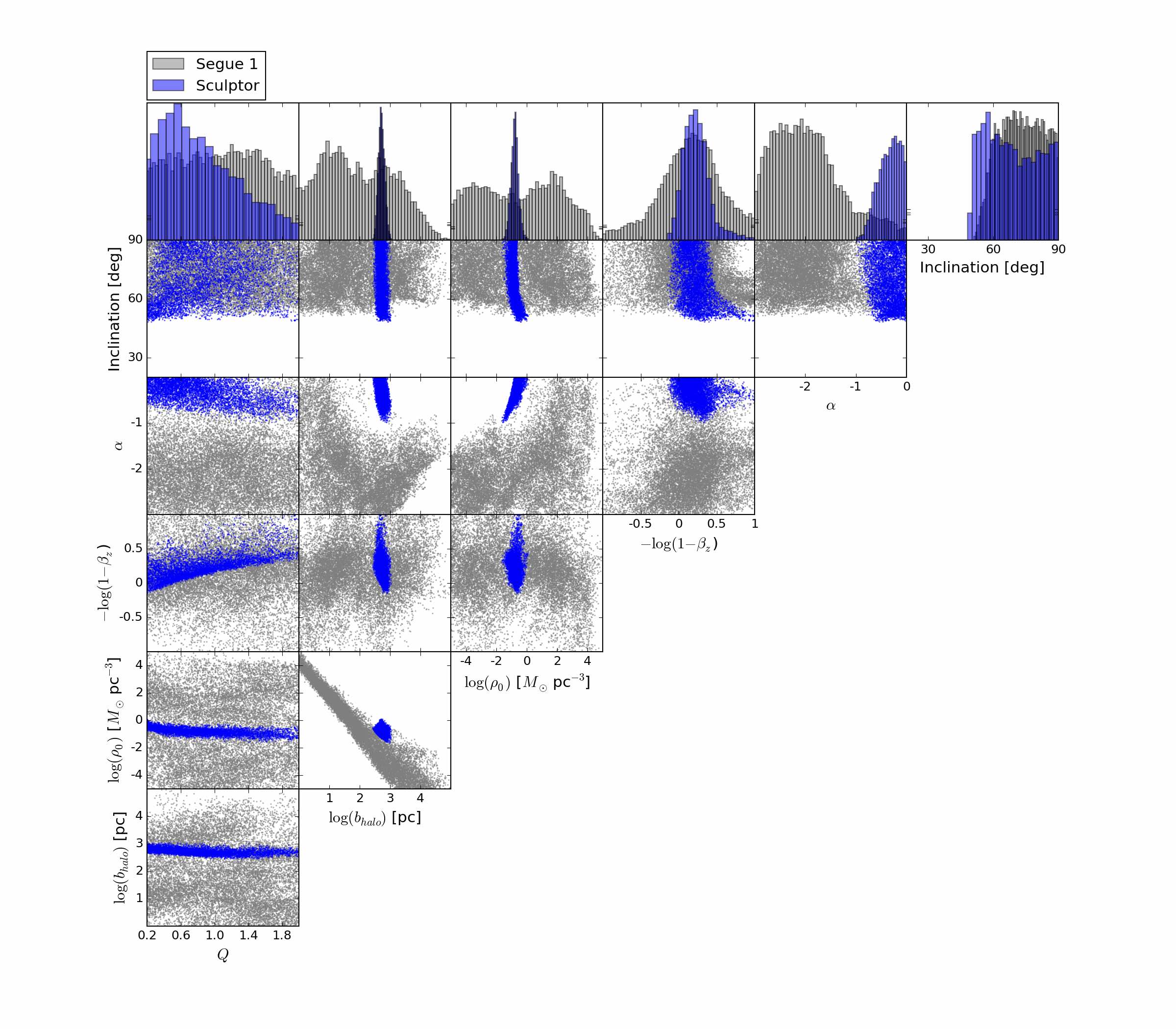}
    \caption{Posterior probability distribution functions of dark matter halo parameters for Sculptor~(blue) and Segue~I~(grey).}
    \label{fig:map}
\end{figure*}
% Figure 3
\begin{figure}
	\includegraphics[scale=0.085]{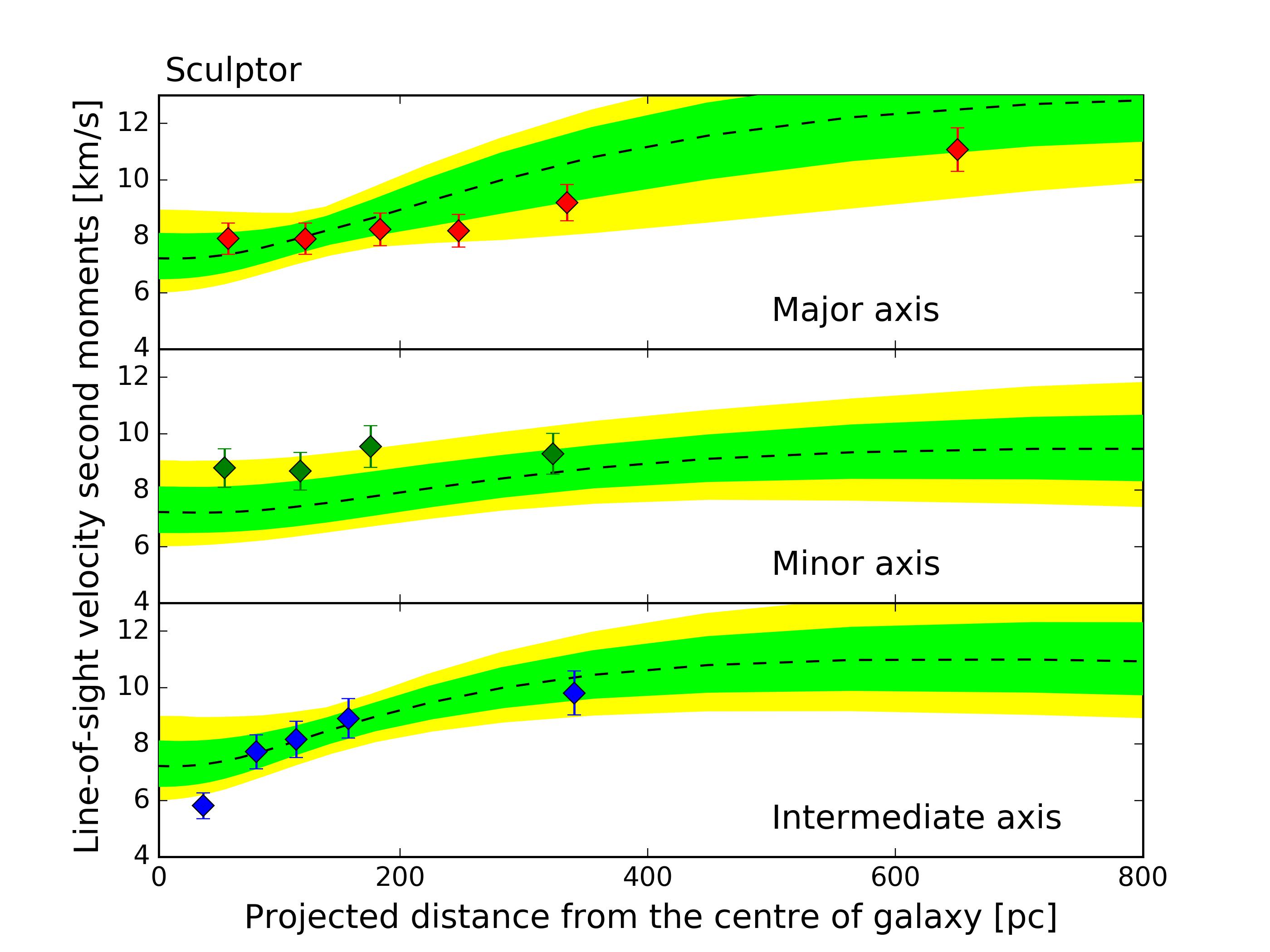}
	\includegraphics[scale=0.085]{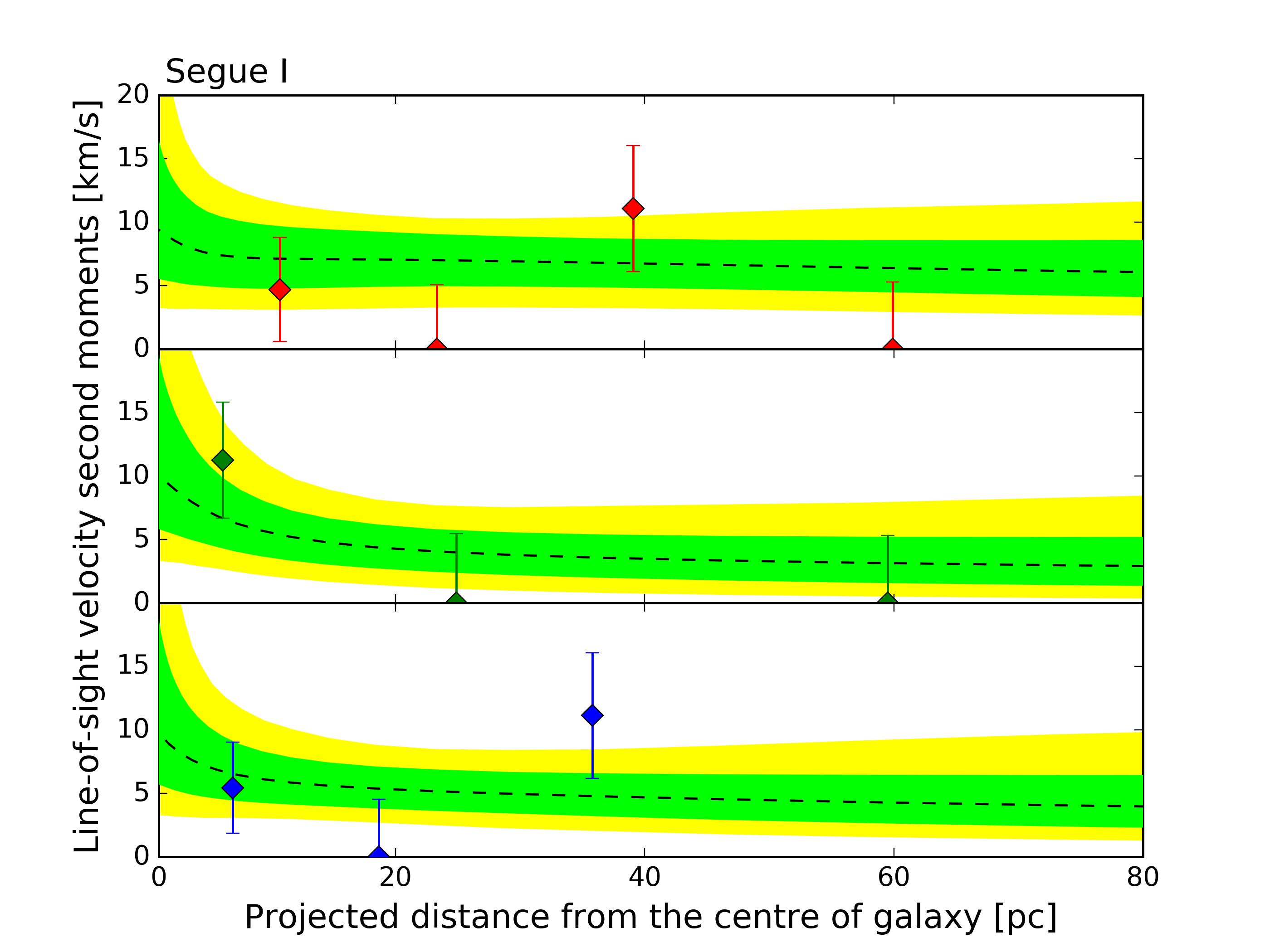}
    \caption{Profiles of line-of-sight velocity second moment along major (first row), minor (second row) and intermediate (third row) axes for Sculptor (top panel) and Segue~I (bottom panel). The colour diamonds with error bars in each panel denote observed second moments. The dashed lines are median second moment values of models and the green and yellow shaded regions encompass the 68 per~cent and 95 per~cent confidence levels from the results of unbinned MCMC analysis. We describe in the main text the methods for generating the binned profiles.}
    \label{fig:bin}
\end{figure}

Besides, the line-of-sight velocity distribution can be influenced by unresolved binary stars, which would significantly inflate the observed velocity second moments. For classical dSphs, the above works in the literature investigated the effect of binary systems on velocity second moments and concluded that the influence of binary systems in these dSphs is, in fact, negligible because their intrinsic velocity second moments are much larger than the velocity distributions inflated by binaries. Although not all of the UFD galaxies were investigated for this effect, some authors considered that a binary star is unlikely to make the measured velocity second moments dramatically inflated~\citep[see][]{SG2007,Simetal2011,Simetal2015,Kopetal2011,Kiretal2013,MC2010}.
Therefore, we suppose that the velocity data of each dSph is not affected by the presence of binary stars.

%%% Sec.5 %%%%%%%%%%%%%%%%%%%%%%%%%%%%%%%%%%%%%%%%%%%%%%%%
\section{Fitting procedure}
Our aim is to obtain the dark matter halo parameters and determine their uncertainties by fitting our mass models to the velocity second moments of each dSph. As described above, the fitting procedure in the current work is different from those in previous axisymmetric works. HC15 fitted their mass models to line-of-sight velocity sec- ond moment profiles built from the individual stellar velocities of dSphs, whilst our work adopts the Gaussian distribution of the line- of-sight velocity to compare the observed and theoretical velocity second moments.

First, we assume that the line-of-sight velocity distribution is Gaussian, centred on the systemic velocity of the galaxy $\langle u \rangle$. Thus we define the likelihood function as follows,
\begin{equation}
{\cal L} = \prod^{N}_{i=1}\frac{1}{(2\pi)^{1/2}[\delta^2_{u,i} + \overline{v^2}(x_i,y_i)]^{1/2}}\exp\Bigl[-\frac{1}{2}\frac{(u_i-\langle u \rangle)^2}{\delta^2_{u,i} + \overline{v^2}(x_i,y_i)} \Bigr],
\end{equation}
where $u_i$ and $\delta_{u,i}$ are the line-of-sight velocity and the observational error of the $i{\rm th}$ star in the available kinematic data set, $(x_i,y_i)$ are the the two-dimensional sky position with respect to the centre of the galaxy, and $\overline{v^2}(x_i,y_i)$ is the theoretical line-of-sight velocity second moment specified by model parameters $(Q,b_{\rm halo},\rho_0,\beta_z,\alpha,i)$ and derived from axisymmetric Jeans equations (see Section~3.1).
The six model parameters are the four parameters of the dark matter halo, and the two parameters of the stellar properties, for which we adopt uniform priors.
The prior ranges of each parameter are

\begin{itemize}
\item $0.1 \leq Q \leq 2.0$;\\
\item $0 \leq \log_{10}[b_{\rm halo}/{\rm pc}] \leq +5$;\\
\item $-5 \leq \log_{10}[\rho_0/(M_{\odot}{\rm pc}^{-3})] \leq +5$;\\
\item $-1 \leq -\log_{10}[1-\beta_z] < +1$;\\
\item $-3 \leq \alpha \leq 0$;\\
\item $\cos^{-1}(q^{\prime}) \leq i \leq 90^{\circ}$.\\
\end{itemize}
As we described in Section~3.1, the lower limit of the inclination angle~$i$ is confined by $q^{\prime 2} -\cos^2i>0$.
Besides the above free parameters, the systemic velocity $\langle u \rangle$ of the system is also included free parameters as a flat prior.
To explore the large parameter space efficiently, we utilize MCMC techniques, based on Bayesian parameter inference, with the standard Metropolis-Hasting algorithm \citep{Metetal1953,Has1970}. 
We take several post-processing steps (burn-in step, the sampling step and length of the chain) to generate independent samples that are insensitive to the initial conditions, and then we obtain the posterior probability distribution function (PDF) of the set of free parameters. 
By calculating the percentiles of these PDFs, we are able to compute credible intervals for each parameter straightforwardly.

% That is, %%% Sec.6 %%%%%%%%%%%%%%%%%%%%%%%%%%%%%%%%%%%%%%%%%%%%%%%%
\section{Results}
In this section, we present the results of the MCMC fitting analysis for seven classical and 17 ultra-faint dSph galaxies with six free parameters, and compute the median and credible interval values from the resulting PDF.

Fig.~\ref{fig:map} displays the posterior PDFs returned by the MCMC procedure for Sculptor and Segue~I. The former is one of the most luminous classical dSphs and the latter is one of the most luminous ultra-faint dSphs. 
Obviously, while the halo parameters $(b_{\rm halo},\rho_0,\alpha)$ for Sculptor dSph are well constrained due to the large size of the kinematic data sample, those for Segue~I are widely distributed in all parameter spaces.
On the other hand, the inclination angles $i$ of both dSphs are also distributed in a wide parameter range and thus, it is difficult to confine the parameter distribution of $i$, which is compatible with the results from previous axisymmetric works~\citep{HC2012,HC2015b}.
In addition, there exists a degeneracy between the shape of the dark halo $Q$ and the stellar velocity anisotropy $\beta_z$, suggested by~\citet{Cap2008} and~HC15, who presented that the variation of these has a similar effect on the line-of-sight velocity second moment profiles.
However, the histogram in the top left panel of Fig.~\ref{fig:map} shows posterior PDF of $Q$, indicating that Sculptor is likely a flattened and oblate dark halo, implying that the best-fitting $Q$ of this galaxy estimated by~HC15 for binned analysis is not deviated very significantly.

Fig.~\ref{fig:bin} displays the line-of-sight velocity second moment profiles along the projected major, minor and intermediate axes for Sculptor and Segue I. To obtain these binned profiles from the observed kinematic data, we utilize the standard technique of using binning profiles. 
First, since we assume an axisymmetric stellar system in this work, we analyse the line-of-sight velocity data by folding the stellar distribution into the first quadrant in each dSph. 
Secondly, we represent the stellar distribution in the sky (of the first quadrant) in two-dimensional polar coordinates $(r,\theta)$, where $\theta=0^{\circ}$ is set along the major axis, and then divide this into three areas in increments of $30^{\circ}$ in the direction from $\theta=0^{\circ}$ to $90^{\circ}$. For convenience, the region between $\theta=$ $0^{\circ}$ and $30^{\circ}$ is regarded as a major axis area, $\theta=30^{\circ}$ - 60$^{\circ}$ as an intermediate axis area, and $\theta=60^{\circ}$ - 90$^{\circ}$ as a minor axis area.
Finally, for each area, we radially separate stars into bins so that a nearly equal number of stars is contained in each bin. The dashed lines and shaded regions denote the median and confidence levels (green: 68 per~cent, yellow: 95 per~cent) of the sampled unbinned MCMC analysis of the model profiles, while the coloured marks with error bars are binned second moments estimated from the observed data. It is found from this figure that our MCMC analysis is in good agreement even with binned profiles. However, we note that since there is a paucity of kinematic sample stars in UFD galaxies, it is difficult to generate binned profiles along three different axes of a dSph. Although Segue~I has a relatively large sample among UFD galaxies, the velocity sample in each bin has around only five stars. Therefore, we are not able to show binned profiles for all UFD galaxies, especially for the galaxies whose kinematic sample is very small.

%%% Table 2 %%%
%%% Table 2 %%%
\begin{table*}
	\centering
	\caption{Parameter constraints for MW dSph satellites. Errors correspond to the $1\sigma$ range of the our analysis.}
	\label{table2}
	\scalebox{0.9}[0.9]{
	\begin{tabular}{cccccccc} % four columns, alignment for each
		\hline\hline
Object    & $Q$ &   $\log_{10}(b_{\rm halo})$[pc]  & $\log_{10}(\rho_0)$ $[$M$_{\odot}$ pc$^{-3}$]       & $-\log_{10}(1-\beta_z)$   & $\alpha$            &$i$ [deg]   &  2$\ln$(BF)$^{*}$\\
		\hline
{\bf Classical dwarfs}     &&&&&&\\
Carina            & $0.6^{+0.6}_{-0.4}$  & $3.5^{+0.7}_{-0.6}$ & $-2.2^{+1.0}_{-0.9}$  & $ 0.2^{+0.2}_{-0.2}$ & $-1.0^{+0.4}_{-0.2}$  & $71^{+14}_{-15 }$  & 24\\
Fornax            & $1.1^{+0.7}_{-0.5}$  & $2.8^{+0.2}_{-0.2}$ & $-1.1^{+0.2}_{-0.2}$  & $ 0.3^{+0.1}_{-0.1}$ & $-0.2^{+0.1}_{-0.2}$  & $72^{+12}_{-15 }$  & 20\\
Sculptor          & $0.8^{+0.6}_{-0.4}$  & $2.7^{+0.1}_{-0.1}$ & $-0.8^{+0.2}_{-0.3}$  & $ 0.2^{+0.2}_{-0.2}$ & $-0.3^{+0.2}_{-0.3}$  & $68^{+16}_{-12 }$  & 22\\
Sextans           & $1.0^{+0.6}_{-0.6}$  & $3.5^{+1.0}_{-0.6}$ & $-2.0^{+0.6}_{-1.1}$  & $ 0.2^{+0.2}_{-0.2}$ & $-0.6^{+0.4}_{-0.5}$  & $72^{+13}_{-13 }$  & 22\\
Draco             & $1.4^{+0.4}_{-0.7}$  & $4.4^{+0.5}_{-0.6}$ & $-2.7^{+0.8}_{-0.6}$  & $ 0.5^{+0.3}_{-0.3}$ & $-1.0^{+0.2}_{-0.1}$  & $59^{+15}_{-7  }$  & 20\\
Leo~I             & $1.0^{+0.7}_{-0.6}$  & $2.8^{+0.9}_{-0.4}$ & $-1.3^{+0.9}_{-1.6}$  & $ 0.0^{+0.2}_{-0.2}$ & $-1.2^{+0.6}_{-0.4}$  & $60^{+19}_{-14 }$  & 16\\
Leo~II            & $1.2^{+0.6}_{-0.7}$  & $3.2^{+0.8}_{-0.7}$ & $-1.7^{+1.0}_{-1.2}$  & $ 0.2^{+0.2}_{-0.2}$ & $-1.0^{+0.5}_{-0.4}$  & $62^{+19}_{-17 }$  & 13\\
%Ursa Minor & No data \\
{\bf Ultra faint dwarfs}    &&&&&&\\
Segue~1           & $1.1^{+0.6}_{-0.6}$  & $1.9^{+1.4}_{-1.1}$ & $-0.4^{+2.7}_{-3.2}$  & $ 0.2^{+0.4}_{-0.4}$ & $-2.0^{+0.8}_{-0.6}$  & $73^{+11}_{-11 }$  & 25\\
Segue~2           & $1.0^{+0.6}_{-0.6}$  & $1.5^{+1.4}_{-1.0}$ & $-3.2^{+2.3}_{-1.2}$  & $-0.3^{+0.5}_{-0.4}$ & $-1.4^{+1.0}_{-1.0}$  & $54^{+23}_{-16 }$  & 12\\
Bo\"otes I        & $1.0^{+0.6}_{-0.6}$  & $3.6^{+0.9}_{-0.9}$ & $-2.4^{+0.8}_{-1.1}$  & $ 0.7^{+0.2}_{-0.2}$ & $-0.7^{+0.5}_{-0.6}$  & $77^{+9 }_{-11 }$  & 16\\
Hercules          & $1.0^{+0.7}_{-0.6}$  & $2.3^{+1.2}_{-1.2}$ & $-1.4^{+2.8}_{-1.9}$  & $ 0.6^{+0.3}_{-0.4}$ & $-1.4^{+1.0}_{-1.1}$  & $82^{+5 }_{-6  }$  & 13\\
Coma~Berenices    & $1.2^{+0.6}_{-0.6}$  & $3.0^{+1.2}_{-0.8}$ & $-0.8^{+0.7}_{-1.5}$  & $-0.3^{+0.6}_{-0.5}$ & $-0.6^{+0.4}_{-0.7}$  & $72^{+12}_{-13 }$  & 15\\
Canes~Venatici~I  & $1.2^{+0.5}_{-0.7}$  & $3.9^{+0.8}_{-0.9}$ & $-2.2^{+0.6}_{-1.0}$  & $ 0.4^{+0.3}_{-0.3}$ & $-0.7^{+0.4}_{-0.5}$  & $74^{+11}_{-11 }$  & 22\\
Canes~Venatici~II & $1.1^{+0.6}_{-0.6}$  & $3.2^{+1.1}_{-0.9}$ & $-1.4^{+1.0}_{-1.6}$  & $-0.3^{+0.6}_{-0.4}$ & $-0.9^{+0.6}_{-0.6}$  & $76^{+10}_{-10 }$  & 12\\
Leo~IV            & $1.1^{+0.6}_{-0.6}$  & $1.6^{+1.8}_{-1.1}$ & $-1.7^{+3.3}_{-2.2}$  & $ 0.0^{+0.6}_{-0.6}$ & $-1.1^{+0.7}_{-1.1}$  & $73^{+11}_{-10 }$  & 12\\
Leo~V             & $1.2^{+0.5}_{-0.7}$  & $3.3^{+1.0}_{-1.5}$ & $-2.6^{+3.6}_{-1.5}$  & $-0.1^{+0.7}_{-0.6}$ & $-1.1^{+0.7}_{-1.1}$  & $78^{+8 }_{-9  }$  & 13\\
Leo~T             & $1.3^{+0.5}_{-0.7}$  & $2.3^{+1.3}_{-0.9}$ & $-0.6^{+1.8}_{-1.8}$  & $-0.3^{+0.4}_{-0.4}$ & $-1.2^{+0.8}_{-0.9}$  & $47^{+26}_{-18 }$  & 12\\
%Willman~I & No data\\
Ursa Major~I      & $0.9^{+0.7}_{-0.5}$  & $2.2^{+1.3}_{-1.4}$ & $-0.8^{+3.5}_{-2.1}$  & $ 0.8^{+0.1}_{-0.2}$ & $-1.5^{+1.0}_{-1.0}$  & $87^{+2 }_{-4  }$ & 16\\
Ursa Major~II     & $1.1^{+0.6}_{-0.6}$  & $2.8^{+1.2}_{-1.0}$ & $-1.2^{+2.5}_{-2.0}$  & $ 0.0^{+0.7}_{-0.6}$ & $-1.5^{+1.0}_{-1.0}$  & $80^{+7 }_{-7  }$ & 13\\
Reticulum~II      & $1.1^{+0.6}_{-0.6}$  & $2.4^{+1.1}_{-1.1}$ & $-0.8^{+1.5}_{-1.4}$  & $ 0.2^{+0.4}_{-0.6}$ & $-1.0^{+0.6}_{-0.7}$  & $79^{+7 }_{-8  }$ & 13\\
Draco~II          & $1.2^{+0.5}_{-0.7}$  & $2.0^{+1.6}_{-1.5}$ & $-1.9^{+2.2}_{-2.2}$  & $-0.2^{+0.5}_{-0.5}$ & $-0.9^{+0.6}_{-1.0}$  & $63^{+19}_{-18 }$ & 13\\
Triangulum~II     & $1.1^{+0.6}_{-0.6}$  & $3.0^{+1.2}_{-1.3}$ & $ 0.3^{+1.0}_{-1.8}$  & $-0.4^{+0.5}_{-0.4}$ & $-0.8^{+0.5}_{-0.8}$  & $64^{+16}_{-17 }$ & 15\\
Hydra~II          & $1.2^{+0.6}_{-0.6}$  & $2.1^{+1.6}_{-1.6}$ & $-3.1^{+1.9}_{-1.4}$  & $-0.3^{+0.5}_{-0.5}$ & $-0.9^{+0.6}_{-1.3}$  & $57^{+22}_{-20 }$ & 13\\
Pisces~II         & $1.1^{+0.6}_{-0.7}$  & $2.4^{+1.4}_{-1.3}$ & $-1.5^{+2.7}_{-2.0}$  & $-0.1^{+0.5}_{-0.6}$ & $-1.4^{+0.9}_{-1.0}$  & $71^{+12}_{-13 }$ & 13\\
	\hline
	\end{tabular}
	}
	\\$^{*}$Bayes factor (BF) is defined by $P(M_{axi}|D)/P(M_{sph}|D)$, where $P(M_{axi}|D)$ and $P(M_{sph}|D)$ are mean posterior probabilities of axisymmetric ($M_{axi}$) and spherical symmetric ($M_{sph}$) mass models, respectively. $D$ indicates observational data for each dSph.
\end{table*}

%%% Table 3 %%%
%%% Table 3 %%%
\begin{table*}
	\centering
	\caption{$J$ and $D$ values for assumed non-spherical models. $J_{0.5}$ and $D_{0.5}$ indicate values integrated within $0.5^{\circ}$, while $J_{\rm max}$ and $D_{\rm max}$ are estimated within $(x_{\rm max}, y_{\rm max})$.}
	\label{table3}
	\begin{tabular}{ccccccc} % four columns, alignment for each
		\hline\hline
		Object & $x_{\rm max}$ &  $y_{\rm max}$ & $\log_{10}[J_{0.5}]$ & $\log_{10}[J_{\rm max}]$  & $\log_{10}[D_{0.5}]$ & $\log_{10}[D_{\rm max}]$ \\
               &  [deg]        &  [deg]         & [GeV$^2$~cm$^{-5}$]            & [GeV$^2$~cm$^{-5}$]                 & [GeV~cm$^{-2}$]                &  [GeV~cm$^{-2}$]                   \\
		\hline
		{\bf Classical dwarfs}     &&&&&&\\
		Carina          & 0.64 & 0.50           & $17.98^{+0.46}_{-0.28}$        & $17.99^{+0.47}_{-0.29}$             & $18.21^{+0.24}_{-0.27}$        & $18.28^{+0.25}_{-0.29}$            \\
		Fornax          & 0.62 & 0.50           & $17.90^{+0.28}_{-0.16}$        & $17.90^{+0.28}_{-0.16}$             & $18.05^{+0.14}_{-0.11}$        & $18.09^{+0.14}_{-0.11}$            \\
		Sculptor        & 0.79 & 0.46           & $18.42^{+0.35}_{-0.17}$        & $18.42^{+0.35}_{-0.17}$             & $18.27^{+0.17}_{-0.13}$        & $18.27^{+0.17}_{-0.13}$            \\
		Sextans         & 0.70 & 1.01           & $17.71^{+0.39}_{-0.21}$        & $17.91^{+0.45}_{-0.30}$             & $18.12^{+0.27}_{-0.21}$        & $18.47^{+0.32}_{-0.27}$            \\
		Draco           & 1.73 & 0.95           & $19.09^{+0.39}_{-0.36}$        & $19.44^{+0.44}_{-0.40}$             & $18.84^{+0.23}_{-0.21}$        & $19.51^{+0.25}_{-0.25}$            \\		
		Leo~I           & 0.18 & 0.11           & $17.45^{+0.43}_{-0.23}$        & $17.45^{+0.42}_{-0.23}$             & $17.10^{+0.25}_{-0.19}$        & $17.10^{+0.25}_{-0.19}$            \\
		Leo~II          & 0.10 & 0.16           & $17.51^{+0.34}_{-0.28}$        & $17.51^{+0.34}_{-0.28}$             & $17.17^{+0.32}_{-0.26}$        & $17.17^{+0.32}_{-0.26}$            \\
		%Ursa Minor & No data \\
		{\bf Ultra faint dwarfs}   &&&&&&\\
		Segue~1         & 0.17 & 0.22           & $17.95^{+0.90}_{-0.98}$        & $17.95^{+0.90}_{-0.98}$             & $17.27^{+0.52}_{-0.51}$        & $17.27^{+0.52}_{-0.51}$            \\
		Segue~2         & 1.10 & 0.12           & $13.09^{+1.85}_{-2.62}$        & $13.09^{+1.85}_{-2.62}$             & $15.20^{+1.17}_{-1.54}$        & $15.20^{+1.17}_{-1.54}$            \\
		Bo\"otes I      & 0.18 & 0.19           & $16.95^{+0.53}_{-0.40}$        & $16.95^{+0.53}_{-0.40}$             & $17.16^{+0.42}_{-0.27}$        & $17.16^{+0.42}_{-0.27}$            \\
		Hercules        & 0.17 & 0.08           & $16.28^{+0.66}_{-0.57}$        & $16.28^{+0.66}_{-0.57}$             & $16.47^{+0.54}_{-0.39}$        & $16.47^{+0.54}_{-0.39}$            \\
        Coma~Berenices    & 0.17 & 0.10         & $18.52^{+0.94}_{-0.74}$        & $18.52^{+0.94}_{-0.74}$             & $17.67^{+0.73}_{-0.53}$        & $17.67^{+0.73}_{-0.53}$            \\ 
		Canes~Venatici~I  & 0.14 & 0.12         & $16.92^{+0.43}_{-0.26}$        & $16.92^{+0.43}_{-0.26}$             & $17.18^{+0.34}_{-0.26}$        & $17.18^{+0.34}_{-0.26}$            \\
		Canes~Venatici~II & 0.12 & 0.03         & $17.23^{+0.84}_{-0.68}$        & $17.23^{+0.84}_{-0.68}$             & $16.71^{+0.62}_{-0.48}$        & $16.71^{+0.62}_{-0.48}$            \\
        Leo~IV            & 0.07 & 0.09         & $15.31^{+1.58}_{-2.90}$        & $15.31^{+1.58}_{-2.90}$             & $15.86^{+0.89}_{-1.50}$        & $15.86^{+0.89}_{-1.50}$            \\
		Leo~V             & 0.18 & 1.52         & $16.24^{+1.26}_{-1.36}$        & $16.24^{+1.26}_{-1.36}$             & $17.37^{+0.60}_{-0.69}$        & $17.37^{+0.60}_{-0.69}$            \\
		Leo~T             & 0.07 & 0.03         & $16.75^{+0.61}_{-0.53}$        & $16.75^{+0.61}_{-0.53}$             & $16.27^{+0.60}_{-0.43}$        & $16.27^{+0.60}_{-0.43}$            \\
		%Willman~I & No data\\
		Ursa Major~I      & 0.17 & 0.08         & $17.48^{+0.42}_{-0.30}$        & $17.48^{+0.42}_{-0.30}$             & $17.02^{+0.44}_{-0.25}$        & $17.02^{+0.44}_{-0.25}$            \\
		Ursa Major~II     & 0.36 & 1.11         & $19.56^{+1.19}_{-1.25}$        & $19.56^{+1.19}_{-1.25}$             & $18.80^{+0.52}_{-0.61}$        & $18.80^{+0.52}_{-0.61}$            \\
        Reticulum~II      & 0.16 & 0.07         & $17.76^{+0.93}_{-0.90}$        & $17.76^{+0.93}_{-0.90}$             & $17.03^{+0.70}_{-0.57}$        & $17.03^{+0.70}_{-0.57}$            \\
		Draco~II          & 0.09 & 0.04         & $15.54^{+3.10}_{-4.07}$        & $15.54^{+3.10}_{-4.07}$             & $15.71^{+1.71}_{-1.98}$        & $15.71^{+1.71}_{-1.98}$            \\ 
		Triangulum~II     & 0.12 & 0.05         & $20.44^{+1.20}_{-1.17}$        & $20.44^{+1.20}_{-1.17}$             & $18.42^{+0.86}_{-0.79}$        & $18.42^{+0.86}_{-0.79}$            \\
		Hydra~II          & 0.08 & 0.04         & $13.26^{+2.12}_{-2.31}$        & $13.26^{+2.12}_{-2.31}$             & $14.69^{+1.18}_{-1.24}$        & $14.69^{+1.18}_{-1.24}$            \\ 
		Pisces~II         & 0.06 & 0.02         & $15.94^{+1.25}_{-1.28}$        & $15.94^{+1.25}_{-1.28}$             & $15.68^{+0.78}_{-0.73}$        & $15.68^{+0.78}_{-0.73}$            \\
	\hline
	\end{tabular}
\end{table*}

Table~\ref{table2} lists the results of the fitting analysis for all dSphs. We show the median and $1\sigma$ (68 per~cent) credible intervals of the free parameters, which correspond to the 50${\rm th}$, 16${\rm th}$ (lower error) and 84${\rm th}$ (upper error) percentiles of the posterior PDFs, respectively.
Column 8 in Table~\ref{table2} shows the Bayes factor, which is the ratio of the mean posterior distribution of the axisymmetric and spherical symmetric mass models.
For the spherical models here, it is assumed that the axial ratios of both dark ($Q$) and bright ($q^{\prime}$) components are unity.
It is clearly found for all of the dSphs that our axisymmetric mass models yield a much better fit than the spherical one.
In the following, using these median and $1\sigma$ error values of the dark halo parameters, we calculate the astrophysical $J$ and $D$~factors and then compare them to values from spherical models and previous works.  

%%% Sec.6.1 %%%%%%%%%%%%%%%%%%%%%%%%%%%%%%%%%%%%%%%%%%%%%%%%
\subsection{Astrophysical factor for annihilation and decay}
\begin{figure*}
	\includegraphics[width=\textwidth]{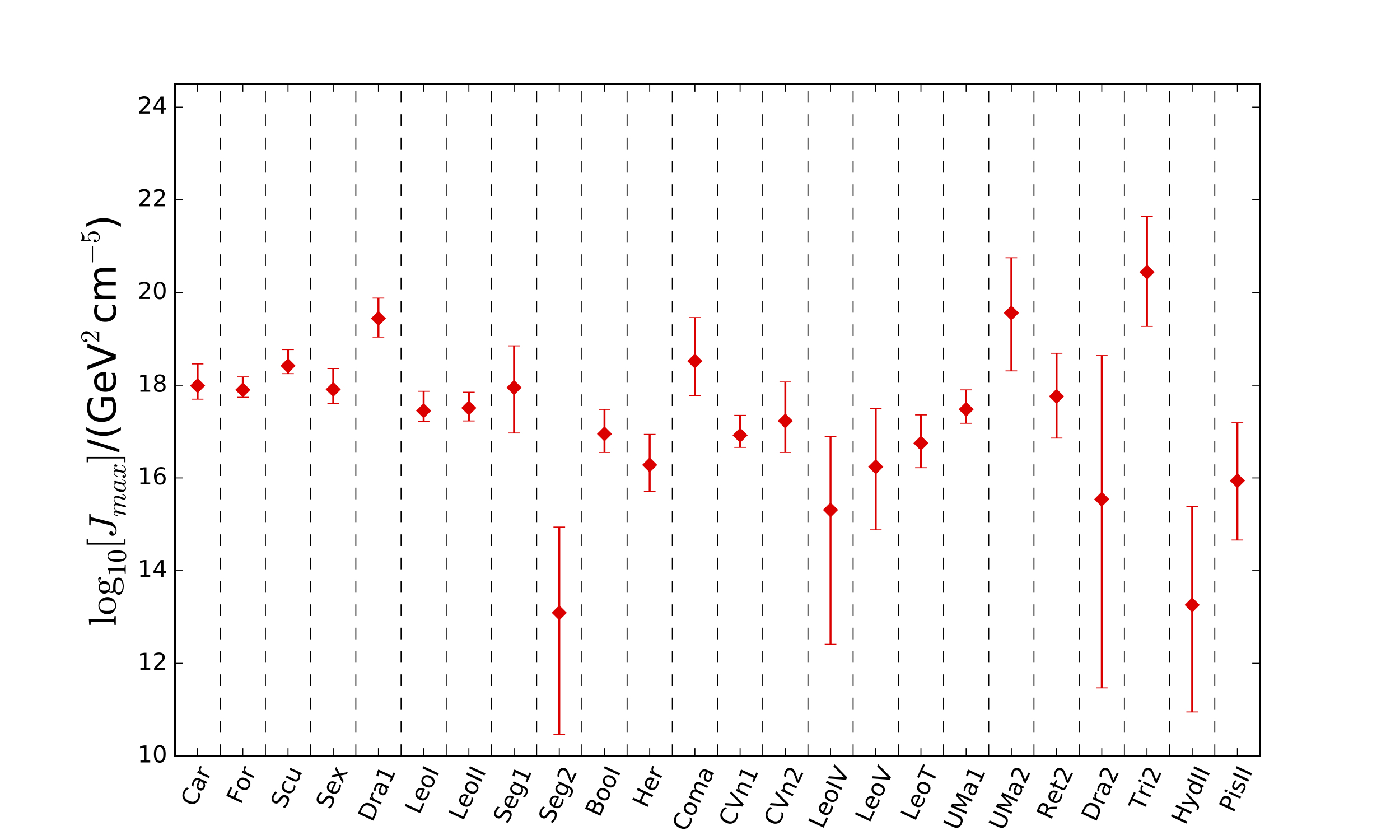}
	\includegraphics[width=\textwidth]{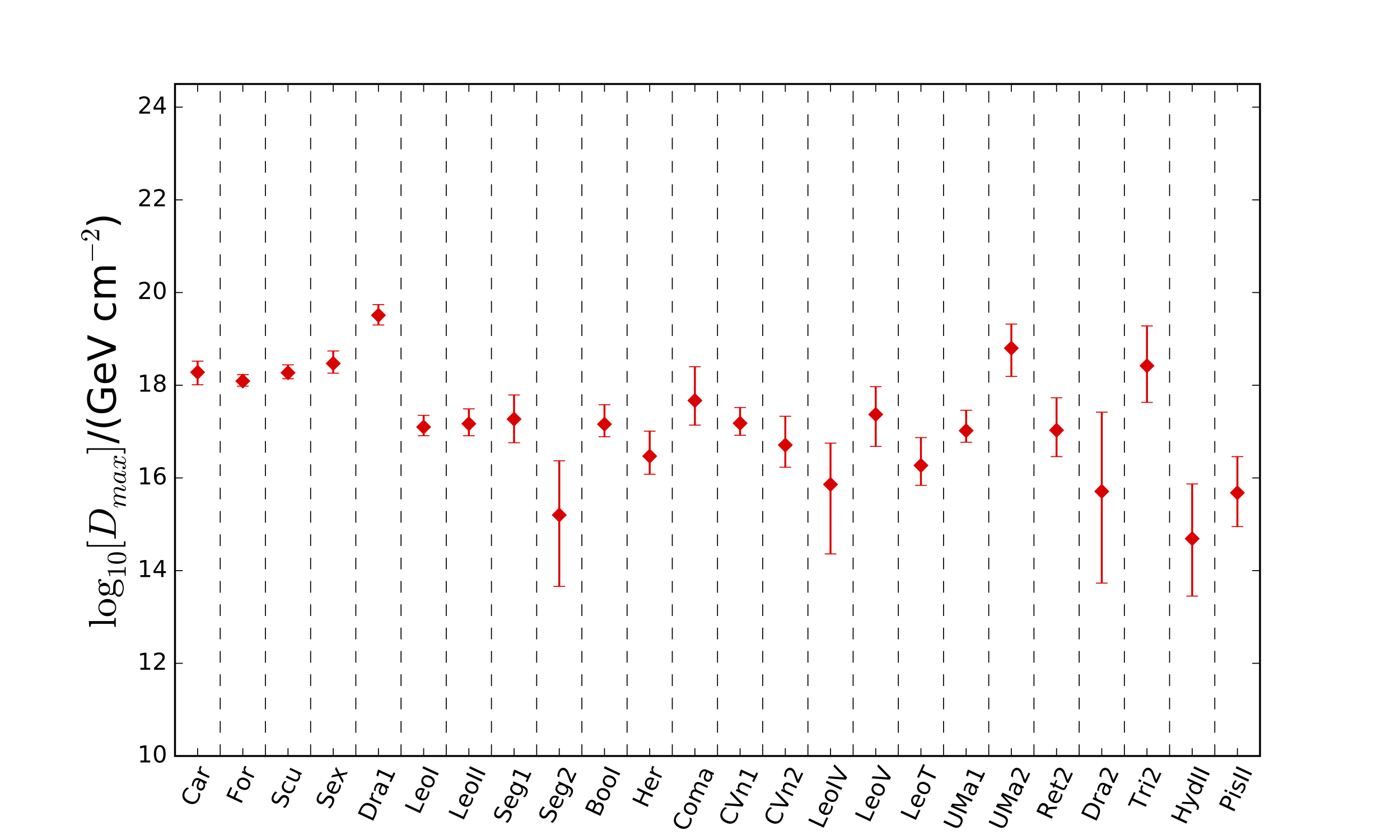}
    \caption{$J$ and $D$ profiles integrated within $(x_{\rm max},y_{\rm max})$ for all the dwarf galaxies. }
    \label{fig:JDmax}
\end{figure*}

The five dark halo parameters $(Q,b_{\rm halo},\rho_0,\alpha,i)$ are used to calculate the dark matter annihilation $J$ factor and decaying dark matter $D$ factor using equations~(\ref{eq:J}) and (\ref{eq:D}).
Table~\ref{table3} shows the $J$ and $D$ values of all dSph's integrated within $0.5^{\circ}$ solid angle and the outermost observed stars as indicated by $(x_{\rm max},y_{\rm max})$. 
If $(x_{\rm max},y_{\rm max})$ are smaller than $0.5^{\circ}$, the astrophysical factors estimated by either integral ranges are identical because we assume that the dark halo extends to the outermost data and is truncated at a corresponding radius.
Fig.~\ref{fig:JDmax} along with Table~\ref{table3} displays the main results of this work.

The top panel of Fig.~\ref{fig:JDmax} shows the estimated $J$ values integrated within $(x_{\rm max},y_{\rm max})$ for all the dSphs.
The error bars of each $J$ value correspond to $1\sigma$ uncertainties based on 16${\rm th}$ and 84${\rm th}$ percentiles of the posterior PDFs.
It is found from Table~\ref{table3} and this figure that among the classical dSphs, Draco has the largest $J$-factor value, while Fornax and Sculptor have the smallest uncertainties of $J$ owing to an abundance of kinematic data. This trend is in agreement with previous studies~\citep[e.g.,][GS15, B15]{Chaetal2011}.
On the other hand, among ultra-faint dSphs, Triangulum~II and Ursa~Major~II are the top and second largest $J$ value, even when including classicals.
However, since their kinematic sample size is limited, $1\sigma$ uncertainties in these values are very much larger than those of all luminous dSphs.
Therefore, all the ultra-faint dSphs, except for Canes~Venatici~I, have pronouncedly larger uncertainties of $J$ than their classical counterparts.

% Figure 4

The bottom panel of Fig.~\ref{fig:JDmax} shows the median $D$ values with $1\sigma$ uncertainties for 24 dSph satellites.
As in top panel of Fig.~\ref{fig:JDmax}, these values are integrated within $(x_{\rm max},y_{\rm max})$.
In contrast to annihilation case, the classical dSphs have systematically higher $D$ values than the UFDs, and the galaxy that has the highest $D$ values is not a UFD galaxy but the classical dSph Draco.
Moreover, Draco has smaller uncertainties of $D$ factor than the other UFD galaxies and, thus, would be the most promising target among the Milky Way satellites. 
The reason why the classicals have higher values is that from equation~(\ref{eq:D}), the $D$ factor largely depends on the extent of the dark halo along the line of sight rather than the inner slope of its density profile, which is actually supported by \citet{Chaetal2011} and GS15.

Some of the UFDs galaxies (e.g. Segue~II and Hydra~II) have significantly low $J$ and $D$ values with large error bars. This is because their kinematic data have such low velocity second moments that their posterior PDFs allow these UFD galaxies to have a ridiculously diffuse and small dark halo, namely having small values of $\rho_0$ and $b_{\rm halo}$ simultaneously.
To remove these anomalous dark halo profiles, some works in the literature considered the tidal radius based on a kinematical and theoretical estimator derived by~\citet{von1957} and \citet{Kin1962} and rejected any dark haloes for which their estimated tidal radius is smaller than the radius of the outermost observed stars~(see GS15 and B15).
However, this estimate for tidal radius is based on several assumptions for simplicity, such as a perfect circular orbit of the satellite and no orbital evolution, thereby implying that the astrophysical $J$ and $D$~factors as well as the tidal radius are affected by large systematic uncertainties due to these assumptions. 
Moreover, we point out that these very low astrophysical values may merely reflect the lack of kinematic data.
Thus, we do not take into account the above prescription in this work, but we present the pristine astrophysical values.

%%% Sec.6.2 %%%%%%%%%%%%%%%%%%%%%%%%%%%%%%%%%%%%%%%%%%%%%%%%
\subsection{Comparison with other works}
\begin{figure*}
	\includegraphics[width=\textwidth]{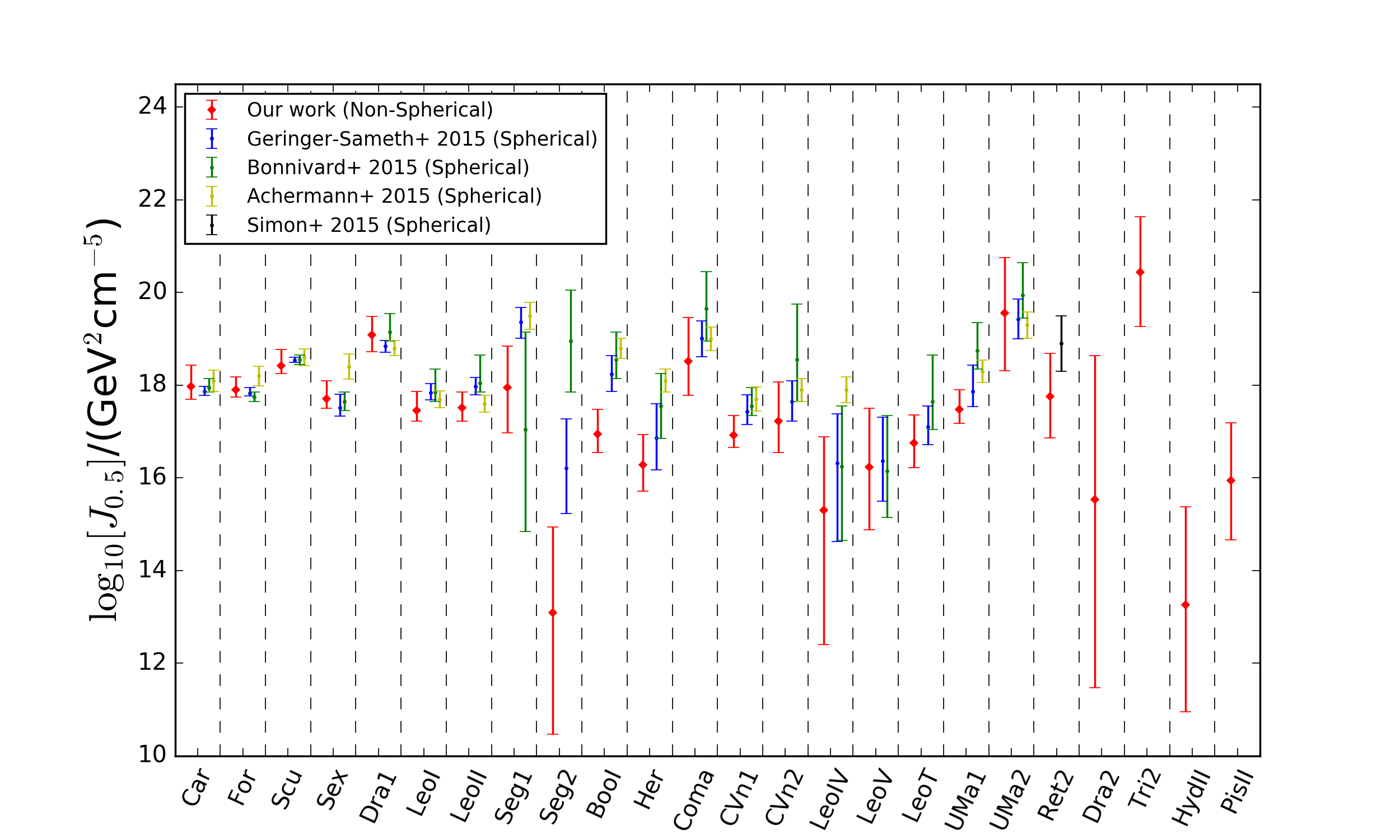}
	\includegraphics[width=\textwidth]{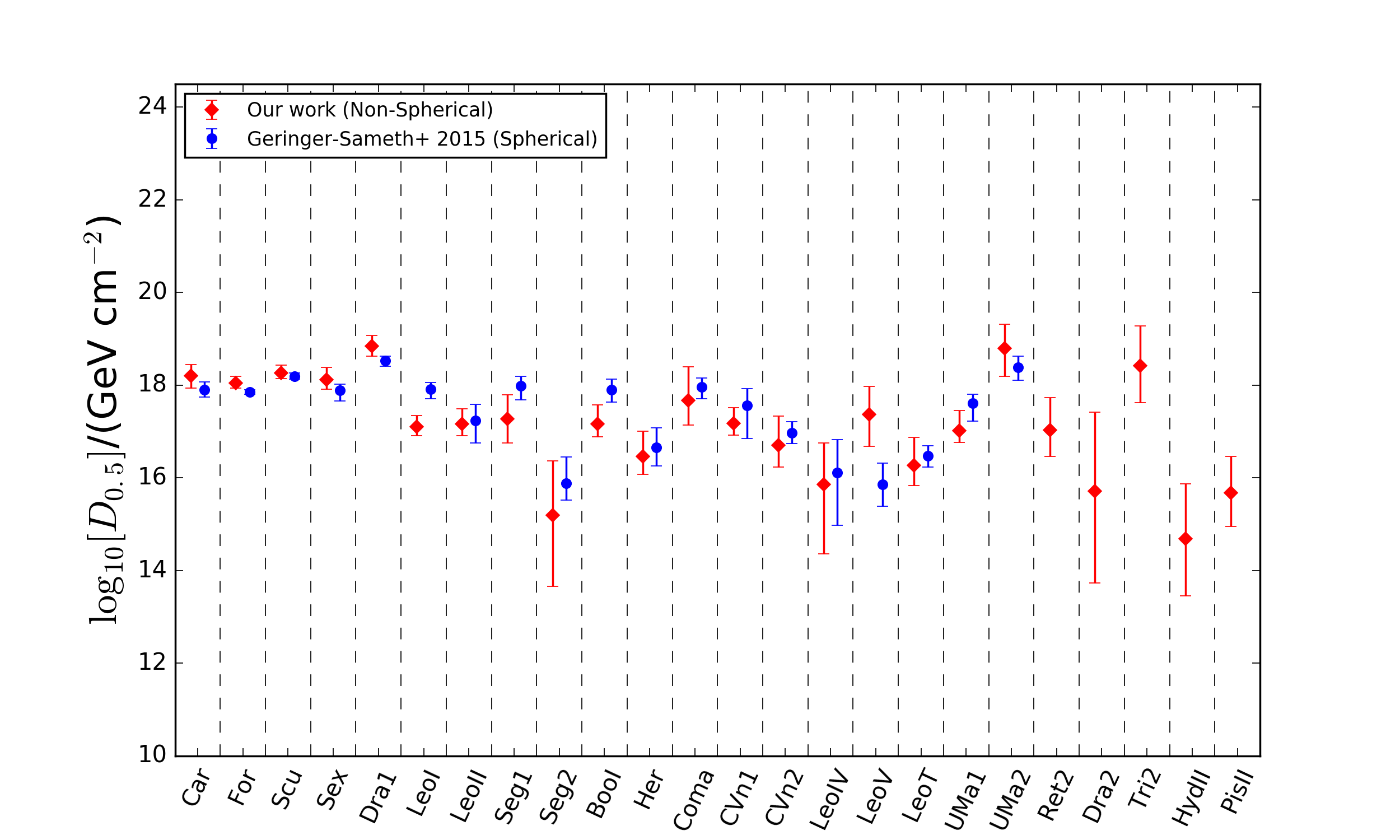}
    \caption{Comparison of $J_{0.5}$~(top) and $D_{0.5}$~(bottom) calculated from axisymmetric and spherical models. The red symbols denote the results of this work. the blue, green, yellow and black ones are estimated by~\citet{Geretal2015}, \citet{Bonetal2015b}, \citet{Acketal2015} and \citet{Simetal2015}, respectively.}
    \label{fig:JD05}
\end{figure*}

In this section, we compare our results in Fig.~\ref{fig:JD05} with other studies based on spherical works.
To do this adequately, we use only $J$ and $D$~factor integrated within a fixed integration angle $0.5^{\circ}$.
Fig.~\ref{fig:JD05} shows a comparison of the $J$ (top) and $D$ (bottom) values of our results with those of previous works.
In this figure, the red diamonds are the median values in this work with $1\sigma$ error bars. The blue, green, yellow and black circles with error bars denote the $J$ values reported by~GS15, B15, \citet{Acketal2015} and \citet[][only Reticulum~2]{Simetal2015}, respectively.
Overall, there are the differences in the median values and the error bar sizes of all dSphs, especially UFDs, between the spherical and non-spherical works.
Naturally, these differences reflect non-sphericity, but that is not all. 
In the following, we describe why these differences occur.

\subsubsection{The difference in the size of the $1\sigma$ uncertainties}
First, the error bars of most dSphs from our estimates are relatively larger than those from other studies. The main reason for this difference is that our axisymmetric analysis fully takes into account the effects of dark halo shape and inclination angle. These systematic uncertainties are consistent with the results of~\citet{Bonetal2015a}, who inspected the influence of the triaxiality of the dark halo on the $J$-factor estimates using mock dSph data taken from {\it The Gaia Challenge.}\footnote{http://astrowiki.ph.surrey.ac.uk/dokuwiki}
B15 (green points) also gives large $1\sigma$ errors of the dSphs except for Carina, Fornax and Sculptor, which have rich data samples.
This is because they adopted not only three dark halo and four velocity anisotropy parameters but also five parameters for the surface brightness profiles of dSphs to construct more flexible mass models.
% Figure 5

On the other hand, the results of GS15~(blue points) are relatively small compared with those of our work and~B15.
They imposed several limitations on dark halo structures, such as the truncation radius and central density intensity required by the cosmological context~(see Section~6 in GS15).
These constraints make an impact on the centered values of the $J$ and $D$~factors as well as the breadth of its uncertainties.

The sizes of the errors obtained by~\citet[][yellow points]{Acketal2015} are also significantly small, even though the UFD galaxies have small data samples.
They assumed NFW dark matter profiles and adopted the \citet{Mar2015} analysis to find a linear relationship between $\log L$ (total luminosity) and $\log V_{\rm max}$ (maximum circular velocity), and between $\log V_{\rm max}$ and $\log r_{\rm max}$ (the radius corresponding to $V_{\rm max}$) of all the dSph satellites, based on the constancy of mass within 300~pc~\citep{Stretal2008}.
That is, they imposed the dark halo properties of classical dSphs on the UFD dark halo, and thus, it is possible that the error bars inferred by~\citet{Acketal2015} for UFD galaxies are artificially suppressed.  
However, \citet{HC2012} argued that the mass of the dSphs enclosed within inner 300~pc varies depending on their total luminosities, in contrast to the conclusion of~\citet{Stretal2008}. Furthermore, recent high-resolution simulations~\citep[e.g.,][]{DiCetal2014a,DiCetal2014b,Onoetal2015} imply that dark halo structures of classical- and UFD-like galaxies are generally different due to the effects of the stellar feedback energy associated with star formation history.
Therefore, we believe that $\log L$, $\log V_{\rm max}$ and $\log r_{\rm max}$ of dSphs do not necessarily have a linear correlation but rather are independent quantities.

\subsubsection{The difference of median values}
Secondly, we focus on the median values of the astrophysical factors.
For most dSphs, these values from our work are systematically smaller than those estimated by others.
This systematic difference can be understood as follows. 
As mentioned in Section~2, the edge of a dark halo in this work is defined by the projected distance of the outermost star $(x_{\rm max},y_{\rm max})$, while~GS15 obtained the de-projected radius of the outermost observed stars using the PDF for this radius, and B15~adopted a theoretically-motivated tidal radius, which is much larger than the outermost data.
Consequently, the truncated distances in our work should be the smallest values, i.e. the $J$ and $D$ factors should be lower than the values in the other works, if the dark haloes considered in these three studies have almost the same properties.

The prior range of dark halo parameters, especially the scale density $\rho_0$, also considerably affect the $J$ and $D$ estimates.
While GS15~and~B15 define the range of dark halo scale density as $-4 \leq \log_{10}[\rho_0/(M_{\odot}{\rm pc}^{-3})] \leq +4$, our work allows the logarithm of scale density to reach $\pm5$, so that a dSph can have a more diffuse or concentrated dark halo than that in previous works.
As seen in the third column of Fig.~\ref{fig:map}, the $\log_{10}(\rho_0)$ posterior PDF of Segue~I favours a diffuse to compact dark halo because the MCMC sample does not distribute as $\log_{10}[\rho_0/(M_{\odot}{\rm pc}^{-3})] > 4$. Thereby, this PDF should be tailed towards a lower scale density, and furthermore, this tendency is commonly found in the other UFD galaxies.

The differences in the inner slope of the dark matter profiles is one of the causes of the differences of the $J$ values, even though this has a low impact on $J$ compared with the above two factors.
GS15~and our study treat the slope of the inner density profile as a free parameter for flexibility, so that dark halos from cored to cusped density profiles can be covered. 
The other works, on the other hand, are confined to cusped density profiles, such as the Einasto and NFW profiles, which make $J$-factor values somewhat large.

In summary, Fig.~\ref{fig:JD05} emphasizes the fact that the median values found for astrophysical factors and their uncertainties strongly depend on the conditions and assumptions.
We believe that our current work is the most reliable estimator for $J$ and $D$ factor as we consider a non-spherical dark halo and the various origins of the uncertainties.

%%% Sec.6.3 %%%%%%%%%%%%%%%%%%%%%%%%%%%%%%%%%%%%%%%%%%%%%%%%
\subsection{Sensitivity line on the $\langle\sigma v\rangle$-$m_{\rm DM}$ map}
By Stacking $J$-factor values of dSph galaxies, we calculate the dark matter annihilation cross-section using 6 yr of Fermi-LAT data (Pass 8). 
The $\gamma$-ray flux with an energy of $E$ from dark matter annihilation in a dSph within a solid
angle $\Delta\Omega$ is given by,
\begin{eqnarray}
\Phi (E, \Delta \Omega) = \left[\frac{\langle \sigma v \rangle}{8 \pi m_{DM}^{2}} \sum_{f} b_{f} \frac{dN_{\gamma}}{dE} \right] \times J_{\Delta \Omega}
\label{eq:flux}
\end{eqnarray}
where $\langle \sigma v \rangle$ is the thermally averaged annihilation cross-section, 
$m_{\rm DM}$ is the mass of a single dark matter particle, 
and $b_f$ and $dN_{\gamma}/dE$ donate the branching fraction of the annihilation 
into the final-state $f$ and the number of photons per energy, respectively. 
The fragmentation function $dN_{\gamma}/dE$ is computed by a Monte Carlo simulation such as Pythia\,\citep{Sjoetal2008} or HERWIG\,\citep{Coretal2001}.
We adopt the function provided by \citet{Ciretal2011}, which includes the non-negligible electro-weak correction\,\citep{Ciaetal2011}.

So far, Fermi-LAT has not found a significant $\gamma$-ray excess originating from dark matter annihilation within the Galactic dSph satellites and, therefore, the background-consistent observation can be used to constrain the dark matter signal flux.
We compute the upper bound on the signal flux by the standard binned Poisson maximum-likelihood method\,\citep{Acketal2015}.
We use the likelihood functions of the signal flux provided by Fermi-LAT\,\citep{Acketal2015}, using the data set of the Pass~8 event selection criteria (P8R2 SOURCE V6) accumulated during 2008 to 2014.
The energy is equally binned logarithmically into $24$ bins within an energy range of $500 {\rm MeV}$-$500 {\rm GeV}$.

In the LAT analysis, the binned likelihood function of the signal flux ${\cal L} (F_{i,j} | D_{i,j})$ is computed independently for each dSph $i$ and each energy bin $j$ under the fixed background flux.
Here $D_{i,j}$ denotes the observed $\gamma$-ray data and $F_{i,j}$ is the signal flux.
The energy spectrum of the signal flux is assumed to be a power-law with a spectral index of two ($dN/dE \propto E^{-2}$),\footnote{
Since the width of the energy bin is small enough, the index of the power law does not significantly affect the final results.} 
while the NFW profile of \citet{Mar2015} is adopted for the spatial dependence of the flux.\footnote{
Fermi-LAT confirms that the choice of the spatial template does not have a significant impact on the analysis\,\citep{Acketal2014}
}
We note that the normalization of $F_{i,j}$ is proportional to $\langle \sigma v \rangle$ and $J$ factor of dSph $i$.

We take the statistical uncertainties on the $J$ factors of each dSph into account by introducing a $J$-factor likelihood term:
\begin{eqnarray}
{\cal L}_J (J_{i}| J_{\text{obs},i}, \sigma_{i}) = \frac{e^{(\log_{10} J_{i} - \log_{10} J_{\text{obs},i})^2/2\sigma_{i}^{2}} 
}{\text{ln}(10) \sqrt{2 \pi \sigma_{i}^{2}} J_{\text{obs}, i}}  .
\end{eqnarray}
Here $J_{i}$ represents the theoretical value of the J factor of dSph $i$, 
$J_{\text obs, i}$ denotes the observed $J$ factor obtained by our fit, and $\sigma_{i}$ is the error of $\log_{10} J_{\text{obs}, i}$.
We adopt the $J$-factor value integrated within an angular radius of $0.5^{\circ}$ shown in Table~\ref{table3}.
Combining the likelihood functions of the signal and the $J$ factor, 
we finally obtain a joint likelihood:
\begin{eqnarray}
\tilde{{\cal L}}(\langle \sigma v \rangle) = \prod_{i,j} 
\underset{J_{i}}{\text{Max}} 
{\cal L} (\langle \sigma v \rangle, J_{i}| D_{i,j})
{\cal L}_{J} (J_{i}| J_{\text{obs},i}, \sigma_{i}) .
\end{eqnarray}
The likelihood $\tilde{\cal{L}}$ is maximized with respect to the nuisance parameter $J_{i}$.
Using this combined likelihood under a fixed $m_{\text{DM}}$, 
we test the hypothesis in which the observed flux contains the dark matter signal with a cross section of $\langle \sigma v \rangle$.
The 95~per~cent confidence level upper limit of the cross-section is computed by solving
$-2 \ln({\cal L}(\langle \sigma v \rangle) /{\cal L} (0))=2.71$.
Here the one-sided chi-square test is applied.

\begin{figure}
	\begin{center}
	\includegraphics[width=\columnwidth]{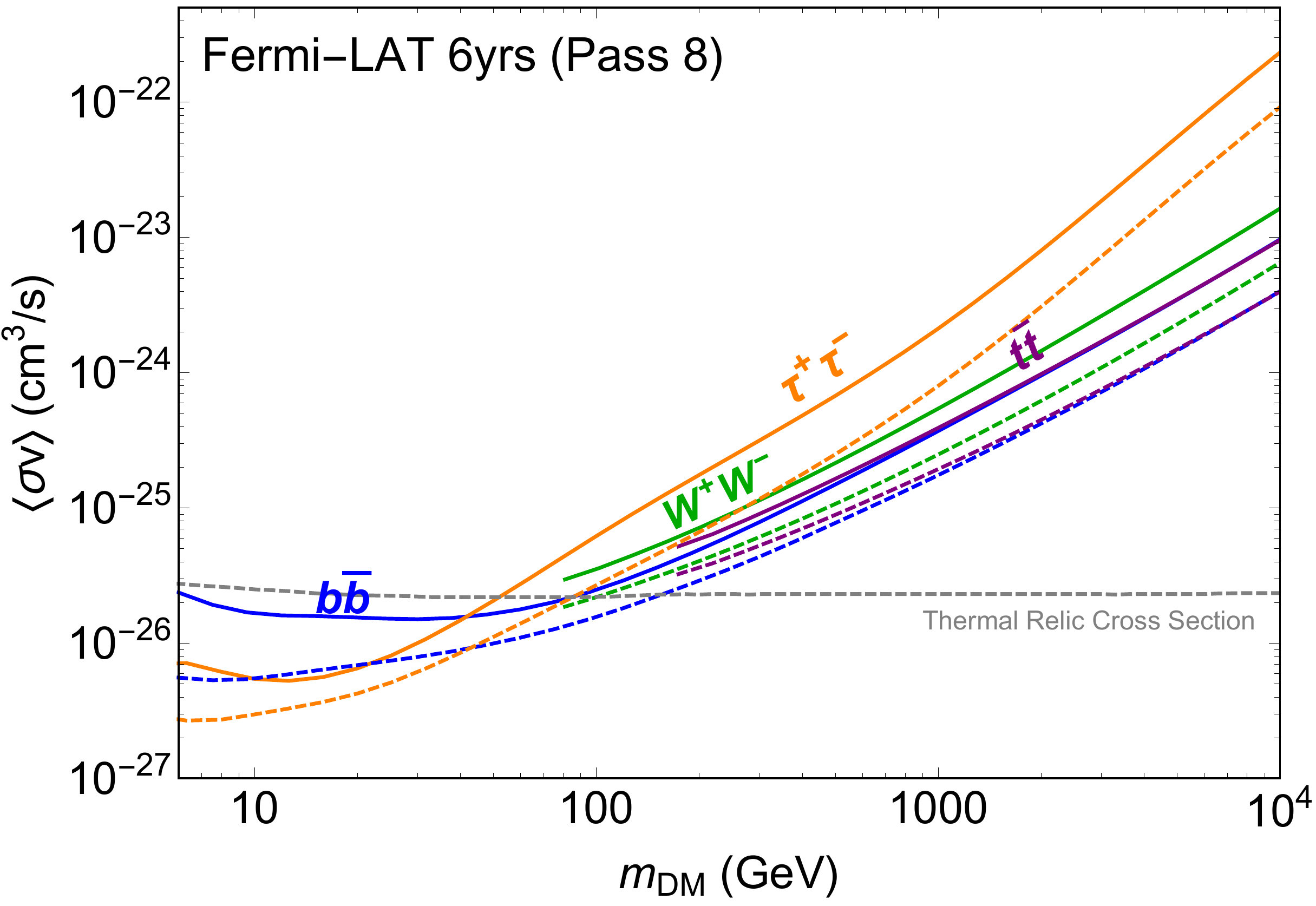}
	\end{center}
    \caption{Comparison of constraints on the dark matter annihilation cross-section estimated from a stacking analysis of 19 dSphs assuming spherical (dashed line) and axisymmetric (solid line) mass models. The blue, purple, green and orange lines denote $b\bar{b}$, $t\bar{t}$, $W^{+}W^{-}$ and $\tau^{+}\tau^{-}$ channels, respectively. The horizontal dashed line is the benchmark value of the thermal relic cross section~\citep{Steetal2012}.}
    \label{fig:cross}
\end{figure}

Fig.~\ref{fig:cross} shows the sensitivity lines of the $b\bar{b}$, $t\bar{t}$, $W^{+}W^{-}$ and $\tau^{+}\tau^{-}$ channels. 
We derive them from a stacking analysis of 19 dSphs excluding eight UFD galaxies~(Segue~II, Leo~V, Leo~T, Reticulum~II, Draco~II, Triangulum~II, Hydra~II, and Pisces~II) to compare fairly between previous spherical and our non-spherical mass models.\footnote{
Since we do not have the kinematical data of Ursa Minor, we do not include it, which improves the LAT sensitivity $\sim 30$ per~cent.
}
It is found from this figure that our analysis with non-sphericity obviously makes each sensitivity line less stringent than the spherical one. 
This is because, as described above, the estimated $J$-factor values in our analysis have large $1\sigma$ errors compared with previous works due to the inclusion of some systematic uncertainties such as non-sphericity, and thus, this is why the constraints on the dark matter annihilation cross-section are relatively weak.

% Figure 6

%%%%%%%%%%%%%%%%%%%%%%%%%%%%%%%%%%%%%%%%%%%%%
Before closing this section, let us discuss the implications
of the present analysis.
First, let us remind ourselves that the most generic $s$-wave cross-section 
of WIMP dark matter, i.e. $\sim 3\times 10^{-26}$cm$^3/$s, 
is one of the primary targets of the indirect searches 
for dark matter.
In particular, the $b\bar{b}$, $W^+W^- (ZZ)$ 
and $t\bar{t}$ channels of this cross-section 
are highly motivated as they are achieved
for neutralino dark matter in the supersymmetric Standard 
Model~\citep{Junetal1996}.
The figure shows that the non-sphericity of the 
dark matter profile leads to constraints about a factor of two weaker 
than the previous constraints.
Accordingly, the constraints on the WIMP mass with a cross-section 
$\sim 3\times 10^{-26}$cm$^3/$s for the $b\bar{b}$ channel
is weakened by about a factor of two.

It should also be emphasized that the indirect searches for dark matter
using $\gamma$-rays are the most important channels in the search 
for the so-called minimal dark matter model~\citep{Ciretal2006,Ciretal2007}. 
In the minimal dark matter model, dark matter fills a single $SU(2)_L$
gauge multiplet and it couples only to $SU(2)_L$ gauge bosons in the Standard Model when it is a fermion. 
As a prominent feature, the annihilation cross-section of dark matter is 
largely enhanced from  $\sim 3\times 10^{-26}$cm$^3/$s
by the so-called Sommerfeld effects~\citep{Hisetal2004,Hisetal2005,Hisetal2006}
in the present Universe, which makes the indirect searches accessible
for a higher dark matter mass region.
In fact, for $SU(2)_L$ triplet fermion dark matter, it has been argued 
that the dark matter mass up to about $3$\,TeV is in tension with 
the $\gamma$-ray observations of the Galactic Centre in Fermi-LAT 
and the HESS telescope~\citep{Cohetal2013,FR2013}.
As the present analysis shows, however, it is important 
to take into account account the systematic uncertainties 
of dark halo evaluations including the effects of non-sphericity 
to draw a final conclusion.

In regard to the $SU(2)_L$ triplet fermion dark matter, let us also emphasize 
that it is also motivated in the so-called anomaly-mediated 
supersymmetry breaking models in the supersymmetric standard model.
There, the $SU(2)_L$ triplet fermion dark matter is naturally achieved 
as the lightest gaugino (the superpartner of the gauge boson) 
and is called the wino. 
After the discovery of the Higgs bosons by the Large Hadron Collider experiments,
the models with anomaly-mediated gaugino mass are considered 
to be one of the most attractive candidates in conjunction 
with the high-scale supersymmetry breaking~\citep{Wells2005,Ibeetal2007,IY2012,IMY2012,HN2012,Haletal2013,NS2014,Arketal2012}. 
This class of models explains the observed Higgs boson mass about $125$\,GeV
~\citep{OYY1991a,OYY1991b,ERZ1991a,ERZ1991b,HH1991} in addition to a good dark matter candidate (i.e. the wino) simultaneously.
To have a better understanding of the systematic uncertainties 
of dark halo evaluations, is quite important to find a hint from  
the fundamental laws of physics such as supersymmetry.

\section{Concluding Remarks}
The Galactic dSphs are ideal targets for constraining particle candidates of dark matter through indirect searches for their annihilations and decays. 
To obtain robust limits on dark matter particle candidates, understanding the true dark matter distribution of these galaxies is of substantial importance. 
In particular, the non-sphericity of the luminous and dark components of these galaxies is one of the major systematic uncertainties of the astrophysical factors for annihilations and decays.
In this paper, by adopting non-spherical mass models developed by HC15, we present non-spherical dark halo structures of seven classical and 17 UFD galaxies and estimate their astrophysical factors. 

In our analysis, Triangulum~II and Ursa~Major~II are the most promising targets for an indirect search of dark matter annihilation, even though they have large uncertainties.
The Draco classical dSph has a $J$ factor only a factor of three lower than those of the above two UFD galaxies but with the very small uncertainties due to the larger number of the sample data. 
For dark matter decay, Draco may be the most detectable and reliable target among all the analyzed dSphs.
Meanwhile, Ursa~Minor classical dSph, which we do not analyse due to not having data, may also be an important object as reported by some works in the literature. 
Thus, we should investigate the dark matter structure in this galaxy and evaluate its astrophysical factors in the near future.
We compare our results for astrophysical factors with other previous studies based on spherical works.
Although the astrophysical factors largely depend on the sample size used for the analysis, the prior range of parameters and the spatial extent of the dark halo, the influence of the non-sphericity of the systems includes the uncertainties of the astrophysical factors as systematics.
We also calculate the dark matter annihilation cross-section using our results and 6 yr of Fermi-LAT data, and compare it with previous work. Since the $1\sigma$ errors of the $J$~factors in our analysis are relatively large because we take into account some systematic uncertainties~(i.e. non-sphericity), the sensitivity lines are more conservative than those of previous spherical works. 
However, we believe that our current work is the most reliable estimator for the astrophysical factor because we consider a non-spherical dark halo and the various origins of their uncertainties, and our axisymmetric models are a much better fit than the spherical ones.

So far, the astrophysical factors of the Galactic dSphs still have large uncertainties because dark halo parameters, especially $Q,\beta_z$ and $i$, are not significantly constrained.
To improve them, we need the photometric and kinematic data over much larger areas as well as a substantial data volume~(see HC15), and adequately removing contaminating stars from the sample data is very important~\citep[e.g.,][]{Ichetal2016}.
The Prime Focus Spectrograph (PFS) attached on the Subaru Telescope~\citep{Sugetal2015} will have remarkable capability to measure kinematic data and metallicities of faint stars in the outer parts of dSphs, and the spectroscopic study of the Galactic dSphs by PFS is one of the primary science goals in the context of the Subaru Strategic Survey~\citep{Taketal2014}.
Therefore, this future spectrograph will allow us to determine robustly dark matter structure in dSphs, thereby enabling to evaluate astrophysical factors without large uncertainties.

\section*{Acknowledgements}
We are grateful to the referee for her/his careful reading of our paper and thoughtful comments.
We would like to give special thanks to Josh Simon and Marla Geha for giving us the kinematic data of UFD galaxies and for useful discussions.
We also thank Evan Kirby and Rosemary Wyse for useful discussions.
This work is supported in part by a Grants-in-Aid for Scientific Research from the Japan Society for the Promotion of Science (JSPS) (No. 25-7047 for M. N. I. and No. 26-3302 for K.H.).
The work of K.I. is also supported by the JSPS Research Fellowships for Young Scientists.
This work has been supported in part by a Grants-in-Aid for Scientific Research from the Ministry of Education, Culture, Sports, Science, and Technology (MEXT), Japan, 
No. 24740151, No. 25105011 (for M. I.), No. 26104009 (for S. M.),  and No. 26287039 (for S. M. and M. I.).
Finally, Kavli IPMU was established is supported by World Premier International Research Center Initiative (WPI), MEXT, Japan.

%%%%%%%%%%%%%%%%%%%%%%%%%%%%%%%%%%%%%%%%%%%%%%%%%%


% Don't change these lines
\bsp	% typesetting comment
\label{lastpage}
\end{document}